\documentclass[preprint,5p,twocolumn]{elsarticle}
\biboptions{sort&compress}

\usepackage{overpic}
\usepackage[hyphens]{url}
\usepackage[hidelinks]{hyperref}
\usepackage{lineno}
\usepackage{optidef}
\usepackage{xcolor}
\usepackage{amsmath}
\usepackage{amssymb}
\usepackage{algpseudocode}
\usepackage{algorithm}
\usepackage{multirow}
\modulolinenumbers[5]
\journal{Applied Energy}
\usepackage{tikz}
\usepackage{pgfplots}
\usetikzlibrary{calc}
\usepackage{bm}
\pgfplotsset{width=7cm,compat=1.14}
\usepgfplotslibrary{statistics}
\usetikzlibrary{positioning}
\usepackage{caption}
\usepackage{subcaption}
\usepackage[utf8]{inputenc}
\usetikzlibrary{arrows}
\usepackage{siunitx}
\usepgfplotslibrary{units}
\pgfplotsset{unit code/.code={\si{#1}}}
\usepackage{diagbox}
\usepackage{multirow}
\newcommand{\vc}[1]{\mathbf{#1}}

\newlength{\separ}
\newlength{\figW}
\newlength{\figH}

\usepackage{colortbl}
\usepackage{xcolor}
\definecolor{green}{RGB}{173,255,47}%
\definecolor{red}{RGB}{255, 0, 0}%
\definecolor{orange}{RGB}{229, 83, 0}%
\definecolor{redPlots}{RGB}{204,37,41}%
\definecolor{greenPlots}{RGB}{62,150,81}%
\definecolor{bluePlots}{rgb}{0.00000,0.44700,0.74100}%
\definecolor{turqoise}{rgb}{0.00000,0.64700,0.64100}%

\usepackage{color}
\definecolor{gray}{rgb}{0.5,0.5,0.5}

\definecolor{dgreen}{rgb}{0,0.5,0}

\newcommand{\nhidn}{n}

\newcommand{\hid}{z}

\newcommand{\rmae}{rMAE}

\begin{document}

\begin{frontmatter}

\title{Forecasting day-ahead electricity prices: A review of state-of-the-art algorithms, best practices and an open-access benchmark
}

\author[delft,ev,vito]{Jesus Lago\corref{cor1}}
\ead{j.lagogarcia@tudelft.nl}

\author[wroclaw]{Grzegorz Marcjasz}
\author[delft]{Bart De Schutter}
\author[wroclaw]{Rafał Weron}

\address[delft]{Delft Center for Systems and Control, Delft University of Technology, Delft, The Netherlands}
\address[ev]{Algorithms, Modeling, and Optimization, Energyville, Genk, Belgium}
\address[vito]{Energy Technology, Flemish Institute for Technological Research (VITO), Mol, Belgium
}
\address[wroclaw]{Department of Operations Research and Business Intelligence, Wrocław University of Science and Technology, Wrocław, Poland
}

\cortext[cor1]{Corresponding author}
\begin{abstract}
While the field of electricity price forecasting has benefited from plenty of contributions in the last two decades, it arguably lacks a rigorous approach to evaluating new predictive algorithms.  The latter are often compared using unique, not publicly available datasets and across too short and limited to one market test samples. The proposed new methods are rarely benchmarked against well established and well performing simpler models, the accuracy metrics are sometimes inadequate and testing the significance of differences in predictive performance is seldom conducted.
Consequently, it is not clear which methods perform well nor what are the best practices when forecasting electricity prices. 
In this paper, we tackle these issues by comparing state-of-the-art statistical and deep learning methods across multiple years and markets, and by putting forward a set of best practices. In addition, we make available the considered datasets, forecasts of the state-of-the-art models, and a specifically designed \texttt{python} toolbox, so that new algorithms can be rigorously evaluated in future studies. 
\end{abstract}


\end{frontmatter}

\section{Introduction}

The increasing penetration of \textit{renewable energy sources} (RES)  in today's power systems makes electricity generation more volatile and the resulting electricity prices harder to predict \cite{BrancucciMartinez-Anido2016,gia:par:pel:16,gro:nan:19,mac:20}. 
On the other hand, advances in \textit{electricity price forecasting} (EPF) constantly provide new tools with the ultimate objective of narrowing the gap between predictions and actual prices. The progress in this field, however, is not steady and easy to follow. 
In particular, as concluded by all major review publications, comparisons between EPF methods  are very difficult since studies use different datasets, different software implementations, and different error measures; the lack of statistical rigor complicates these analyses even further \cite{Weron2014,Nowotarski2018,zie:ste:18,hon:etal:20}. In particular: 
\begin{itemize}
    \item There are several studies comparing \textit{machine learning} (ML) and statistical methods but the conclusions of these studies are contradictory. Typically, studies considering advanced statistical techniques only compare them with simple ML methods \cite{Uniejewski2018a,Marcjasz2019,Cruz2011} and show that statistical methods are obviously better. Conversely, studies proposing new ML methods only compare them with simple statistical methods \cite{Wang2016b,Ugurlu2018,Zhang2018,Luo2019,Chen2019} and show that ML models are more accurate.

    \item In many of the existing studies \cite{Chang2019,Gao2019,Nazar2018,Zhou2018,Singh2018,Yang2017,Chinnathambi2018} the testing periods are usually too short to yield conclusive results. In some cases, the test datasets are limited to one-week periods \cite{Olamaee2016,Darudi2015,Yang2017,Ghayekhloo2019,AruldossAlbertVictoire2018,Zahid2019,Jiang2018,Zhou2019}; this ignores the problem of special days, e.g.\ holidays, and is not  representative for the    performance of the proposed algorithms across a whole year. As argued in \cite{Weron2014}, to have meaningful conclusions, the test dataset should span at least a year.
    
    \item Some of the existing papers do not provide enough details to reproduce the research. The three most common issues are: (i) not specifying the exact split between the training and test dataset \cite{Aggarwal2017,Hong2014a,Talari2017,Singh2017,Khan2017,Afrasiabi2019a,Zhu2018}, (ii) not indicating the inputs used for the prediction model \cite{Wang2017a,Khan2017,Shrivastava2014,Jiang2015,Afrasiabi2019a}, and (iii) not specifying the dataset employed  \cite{Bento2018,Khajeh2017,Singh2018,Talari2017}. 
    This obviously prevents other researchers from validating the research results.
\end{itemize}

These three problems have aggravated over the last years with the increase in popularity of \textit{deep learning} (DL). While new published papers on DL for EPF appear almost every month, and most claim to develop models that obtain state-of-the-art accuracy, the comparisons performed in those papers are very limited. Particularly, the new DL methods are usually compared with simpler ML methods \cite{Lago2018,Kuo2018,Zhou2019,Zahid2019,Mujeeb2019,Atef2019,Mujeeb2019a}. This is obviously problematic as such comparisons are not fair. Moreover, as the proposed methods are not compared with other DL algorithms, new DL methods are continuously being proposed but it is unclear how the different models perform relatively to each other.

Similar problems arise in the context of \textit{hybrid methods}. In recent years, very complex hybrid methods have been proposed. Typically, these hybrid models are based on combining a decomposition technique, a feature selection method, an ML regression model, and sometimes a type of genetic algorithm for optimization purposes. As with DL algorithms, these studies usually avoid comparisons with well-established methods \cite{Khajeh2017,Singh2018,Lahmiri2017,Peter2016,Singh2017,Darudi2015,Naz2019} or resort to comparisons using outdated methodologies \cite{Ghayekhloo2019,Anamika2018,Gao2018,Yang2017,Bento2018,Olamaee2016,Zhu2018}. In addition, while a specific genetic algorithm or decomposition technique is considered, most of the studies do not analyze the effect of selecting a variant of these techniques \cite{Olamaee2016,Singh2018,Naz2019,Anamika2018,Gao2018}. Thus, the relative importance of each of the different components of the hybrid methods it is not even clear.

\subsection{Motivation and contributions}

The above mentioned problems call for three actions.

Firstly, implementing in a popular programming environment (e.g.\ \texttt{python}), thoroughly testing and making available \textit{a set of simple but powerful open-source forecasting methods}, which can potentially obtain state-of-the-art results, and that researchers can easily use to evaluate any new forecasting model. 

Secondly, collecting and making freely available to the EPF community \textit{a set of representative benchmark datasets} that researchers can use to evaluate and compare their methods using long testing periods. Although, some datasets are available for download without restrictions, e.g.\ as supplements to published articles \cite{Hong2016} or sample transaction data \cite{noordpool}, they are typically limited in scope (one market, a 2-3 year timespan or price series only). Hence, conclusions from such datasets are limited, results can hardly be extrapolated to other markets, and the relevance of the studies using such data are not entirely clear.

Thirdly, \textit{putting forward a set of best practices} so that the conclusions of EPF studies become more meaningful and fair comparisons can be made.





In this paper, we try to tackle the above via three distinct contributions:
\begin{enumerate}
	\item We analyze the existing literature and select what could arguably be considered as state-of-the-art among statistical and machine learning methods: the \textit{Lasso Estimated  AutoRegressive} (LEAR) model\footnote{Originally introduced in \cite{Uniejewski2016} under the name \textit{LassoX} and based on the \textit{fARX} model, a  parameter-rich autoregressive specification with exogenous variables. The name refers to the \textit{least absolute shrinkage and selection operator} (LASSO) \cite{Tibshirani1996} used to jointly select features and estimate their parameters. Very similar models were used in \cite{Ziel2018} under the name \textit{24lasso}$_{DoW,nl}$ and in \cite{Uniejewski2018} under the name \textit{24Lasso}$_{\textit{1}}$.} \cite{Uniejewski2016} and the \textit{Deep Neural Network} (DNN) \cite{Lago2018a}, a relatively simple and automated DL method that optimizes hyperparameters and features using Bayesian optimization.
	Then, we make our models open-source and available to other researchers as part of an open-source \texttt{python} library \url{https://github.com/jeslago/epftoolbox} specially designed for this study to provide a common research framework for EPF research \cite{benchmarkwebsite}. Besides the models, we also provide extensive documentation \cite{epftoolboxdoc} for the library.
	
	\item We propose a set of five open-access benchmark datasets spanning six years each, that represent a range of well-established day-ahead, auction type power markets from around the globe. The datasets contain day-ahead electricity prices at an hourly frequency and two relevant exogenous variables each. They can be accessed from the mentioned \texttt{python} library \cite{benchmarkwebsite} that is specially designed for this study. Together with the datasets, the library also includes the forecasts of the open-access methods across the five benchmark datasets so that researchers can quickly make further comparisons without having to re-train or re-estimate the models.

	\item We provide a set of best practice guidelines to conduct research in EPF so that new studies are more sound, reproducible, and the obtained conclusions are stronger. In addition, we include some of the guidelines, e.g.\ adequate evaluation metrics or statistical tests, in the the mentioned \texttt{python} library \cite{benchmarkwebsite} that is specially designed for this study to provide a common research framework for EPF research
\end{enumerate}

\subsection{Paper structure}

The remainder of the paper is organized a follows. Section \ref{sec:litreview} performs a literature review of the current state of EPF. Sections \ref{sec:ben} and \ref{sec:mod} respectively present the open-access benchmark datasets and the open-source benchmark models. Section \ref{sec:guidelines} describes the set of guidelines and best practices when performing research in EPF. Section \ref{sec:eval} discusses the forecasting results for all five datasets. Finally, Section \ref{sec:check} provides a summary and a checklist of the requirements for meaningful EPF research.

\section{Literature review}
\label{sec:litreview}
\begin{figure}[htb]
	\centering
	\includegraphics[width=.8\columnwidth]{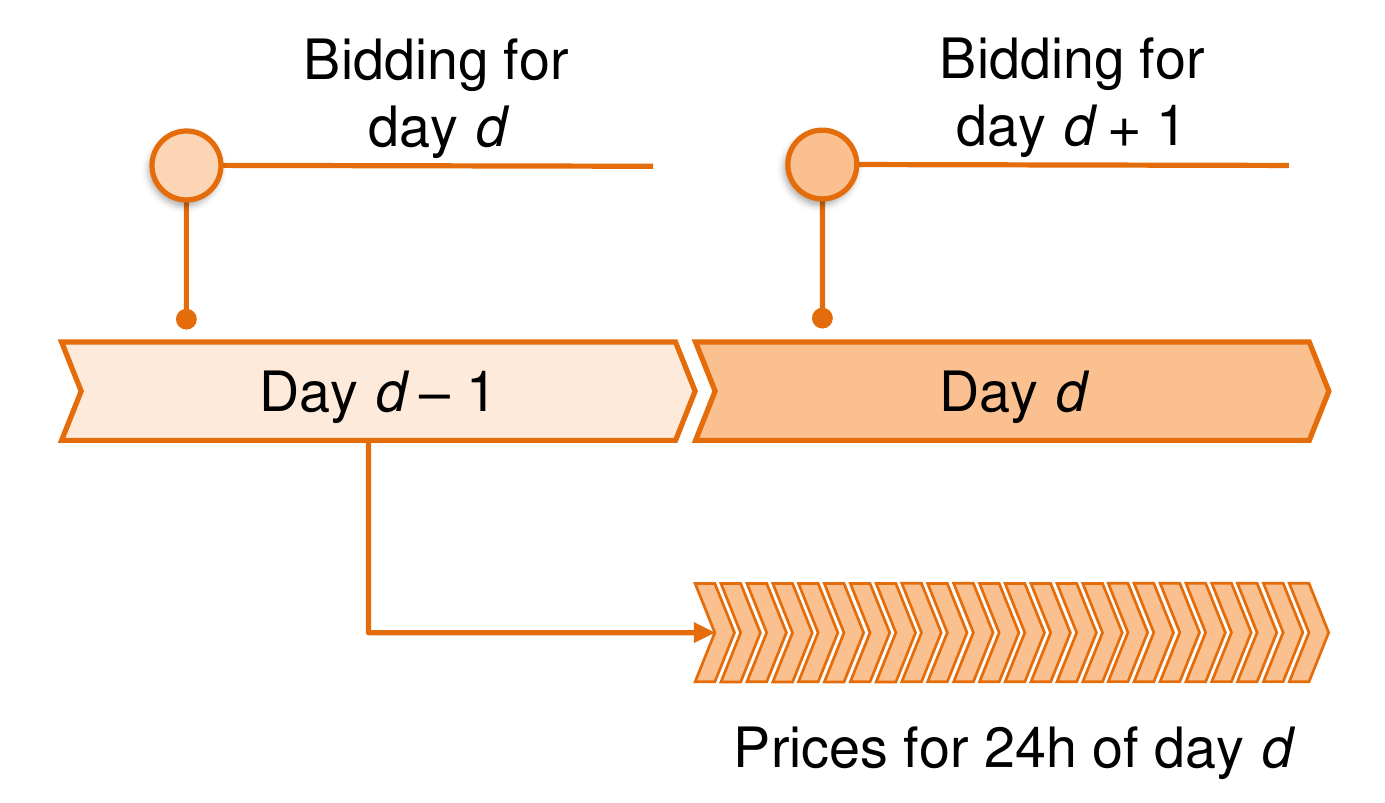}
	\caption{Illustration of the \textit{day-ahead} auction market, where wholesale sellers and buyers submit their bids before gate closure on day $d-1$ for the delivery of electricity during day $d$; the 24 hourly prices for day $d$ are set simultaneously, typically around midday.}
	\label{fig:epf_DA_bidding}
\end{figure}

The field of EPF aims at predicting the spot and forward prices in wholesale markets, either in a point or probabilistic setting. However, given the diversity of trading regulations available across the globe, EPF always has to be tailored to the specific market. For instance, the workhorse of European short-term power trading is the \emph{day-ahead} market with its once-per-day uniform-price auction, see Fig. \ref{fig:epf_DA_bidding}. On the other hand, the Australian National Electricity Market operates as a real-time power pool, where a dispatch price is determined every five minutes and six dispatch prices are averaged every half hour as pool prices \cite{may:tru:18}, while electricity forward markets share many aspects with those of other energy  commodities (oil, gas, coal), and quite often are only financially settled \cite{aid:15}.  

As the field of EPF is very diverse, a complete literature review is out of the scope of this paper. Instead, this section is intended to provide an overview of the the three families of methods, i.e.\ statistical, ML, and hybrid methods, proposed for point forecasting in day-ahead markets since 2014, i.e.\ since the last comprehensive literature review of Weron \cite{Weron2014}. The more recent reviews either focused on short-terk \cite{Nowotarski2018} and medium-/long-term \cite{zie:ste:18} probabilistic EPF, were not that comprehensive in scope \cite{mac:wer:19,wer:zie:20}, or concerned electricity derivatives \cite{aid:15}.
Furthermore, our survey puts a special emphasis on DL and hybrid methods as this is the area of EPF characterized by the most rapid development and, at the same time,  troubled by non-rigorous empirical studies which motivated us to write this paper in the first place.

\subsection{Statistical methods}

Most models in this class rely on linear regression and represent the dependent (or output) variable, i.e.\ the price $p_{d,h}$ for day $d$ and hour $h$, by a linear combination of independent (or predictor, explanatory) variables, also called regressors, inputs, or features:
\begin{equation}\label{eqn:regression}
p_{d,h} = \boldsymbol{\theta}_h \mathbf{X}_{d,h} + \varepsilon_{d,h},
\end{equation}
where $\boldsymbol{\theta}_h=[\theta_{h,0}, \theta_{h,1}, ..., \theta_{h,n}]$ is a row vector of coefficients specific to hour $h$, $\mathbf{X}_{d,h}=[1, X_{d,h}^1, ..., X_{d,h}^n]\top$ is a column vector of inputs and $\varepsilon_{d,h}$ is an error term; the intercept $\theta_{h,0}$ can be set to zero if the data is demeaned beforehand. Note that here we are using a notation common in day-ahead forecasting, which emphasizes the vector structure of these price series, see Fig. \ref{fig:epf_DA_bidding}. Alternatively we could use single indexing: $p_{t}$ with $t=24d+h$. Although the multivariate modeling framework has been shown to be marginally more accurate than the univariate counterpart, both approaches have their pros and cons \cite{Ziel2018,gia:rav:ros:20}.

In the last few years, there have been several key contributions in the field of statistical methods for EPF. Arguably, the most relevant of them has been the appearance of linear regression models with a large number of input features that utilize regularization techniques \cite{Tibshirani1996,Zou2005}. Classically, the regression model in \eqref{eqn:regression} is estimated using \textit{ordinary least squares} (OLS) by minimizing the \textit{residual sum of squares} (RSS), i.e.\ squared differences between the predicted and actual values. However, if the number of regressors is large, using the \textit{least absolute shrinkage and selection operator} (LASSO) \cite{Tibshirani1996} or its generalization the \textit{elastic net} \cite{Zou2005} as implicit feature selection methods have been shown to improved the forecating results \cite{Ziel2015,Uniejewski2016,Ziel2016,Ziel2018,Uniejewski2018,Lago2018a}. In particular, by jointly minimizing the RSS and a penalty factor of the model parameters (see Section \ref{sec:farxmodel} for details), these two implicit regularization techniques set some of the parameters to zero and thus effectively eliminate redundant regressors. As shown in the cited studies \cite{Ziel2015,Uniejewski2016,Ziel2016,Ziel2018,Uniejewski2018,Lago2018a}, these parameter-rich\footnote{We define a parameter-rich linear model as a model with multiple regressors.} regularized regression models exhibit superior performance.
It is important to note that such an approach, called here \textit{Lasso Estimated AutoRegressive} (LEAR), is in fact hybrid since LASSO (and electic nets) are considered ML techniques by some authors. However, we classify it as statistical because the underlying model is autoregressive. 

Aside from proposing parameter-rich models and advanced estimators, researchers have also improved the field by considering a variety of additional preprocessing techniques. Most notably, models using so-called variance stabilizing transformations \cite{sch:11,dia:pla:16,Uniejewski2018a,Ziel2018} and long-term seasonal components \cite{now:tom:wer:13,Nowotarski2016,lis:pel:18,mar:uni:wer:19} have been proposed and shown to result in statistically significant improvements. However, the applicability of these two techniques varies greatly: due to very common occurrence of price spikes, variance stabilizing transformations have become a standard and replaced the commonly used logarithmic transformation (no longer applicable due to zeros and negative values\footnote{The logarithmic of 0 or a negative value is undefined.}) to normalize electricity prices. By contrast, the applicability of long-term seasonal components has been more limited and it is unknown whether their beneficial effect is limited to relatively parsimonious regression models or also holds for parameter-rich models.

A third innovation in the field is an ensemble, i.e.\ a method that combines individual forecasting models to enhance the accuracy,  that combines multiple forecasts of the same model calibrated on different windows. In this context, two different studies \cite{Hubicka2019,Marcjasz2018} showed that the best results are obtained with a combination of a few short (spanning 1-4 months) and a few long calibration windows (of approximately two years). Said ensembles were able to significantly outperform predictions obtained for the best ex-post selected calibration window \cite{Hubicka2019,Marcjasz2018}. But again, it has not been shown to date whether this effect is limited to relatively parsimonious regression models or also holds for LEAR models.


\subsection{Deep learning}
 In the last five years, a total of 28 deep learning papers in the context of EPF have been published\footnote{This data is based on two searches in Scopus looking for keywords in the title, abstract, and keywords. The first search is based on the following query  \texttt{TITLE-ABS-KEY(((("forecasting electricity") OR ("predicting electricity"))  AND  (("electricity spot") OR ("electricity  day-ahead")  OR  ("electricity  price"))) OR ((("price  forecasting")  OR  ("price  prediction") OR ("forecasting  price")  OR  ("predicting  price") OR ("forecasting  spikes")  OR  ("forecasting  VAR")) AND  (("electricity  spot  price")  OR  ("electricity price")  OR  ("electricity  market")  OR  ("day-ahead market") OR ("power market"))) AND ("deep") AND ("learning"))}. The second search is very similar but replacing \texttt{("deep") AND ("learning")} by \texttt{("neural") AND ("network")}.}. Moreover, this number has been steadily increasing: while in 2016 there was only one paper and in 2017 none, in 2018 there were 11, and in 2019 there were 16. Despite this trend, most of the published studies are very limited: the comparisons are too simplistic, e.g.\ avoid state-of-the-art statistical methods, and their results cannot be generalized. 

The first published DL paper \cite{Wang2016b} proposes a deep learning network using stacked denoising autoencoders. The paper, despite being the first, provides a better evaluation than most studies: the new method is compared not only against machine learning techniques but also against two statistical methods. Yet, the evaluation is limited as it is done considering three months of test data and employing simple models for comparison. In the second published DL paper \cite{Lago2018}, a DNN for modeling market integration is proposed. While the method is evaluated over a year of data, the study is also limited as the proposed model is not compared against other machine learning or statistical methods. 

In the third published paper \cite{Lago2018a}, four DL models (a DNNs, two \textit{recurrent neural networks (RNNs)}, and a \textit{convolutional network (CNN)}) are proposed. This study is, to the best of our knowledge, the most complete study up to date. In particular, the proposed DL models are compared using a whole year of data against a benchmark of 23 different models, including 7 machine learning models, 15 statistical methods, a commercial software. Moreover, among the statistical methods, the comparison includes the fARX-Lasso and fARX-EN, i.e.\ the state-of-the-art statistical methods. While the study shows the superiority of the DL algorithms,  very strong conclusions are not possible as the study only considers a single market.

The studies that followed in 2018 focused on one of three topics: 1) evaluating the performance of different deep recurrent networks \cite{Ugurlu2018,Chinnathambi2018,Mujeeb2018,Zhu2018}; 2) proposing new hybrid methods based on CNNs and LSTMs \cite{Kuo2018,Zhang2018,Xie2018,Ugurlu2018a}; or 3) employing regular DNN models \cite{Chinnathambi2018}. Independently of the focus, they  were all more limited than the first and the third studies \cite{Wang2016b,Lago2018a} as they failed to compare the new DL models with state-of-the-art statistical methods and/or to employ long enough datasets to derive strong conclusions.

In detail, \cite{Ugurlu2018} studies the use of RNNs for forecasting electricity prices but the comparison is done in a single market and against simple statistical methods (a seasonal \textit{auto regressive integrated moving average} (ARIMA) model, a Markov regime-switching model, and a self exciting threshold model). Moreover, while the comparison includes other DL methods, it avoids comparison with simpler ML techniques. Ref.\ \cite{Kuo2018} proposes a hybrid DL methods composed of a CNN and a \textit{long short-term memory} (LSTM) (a type of recurrent network) for forecasting balancing prices. However, the new model is only compared against simple ML benchmarks and the evaluation is done using different periods comprising three months for training and 1 month for testing. Similarly, \cite{Zhang2018} proposes another hybrid model combining a CNN and an LSTM, but the model is only compared against two naive statistical methods: an \textit{auto regressive moving average} (ARMA) and a \textit{generalized autoregressive conditional heteroskedasticity} (GARCH)   model.
 
 In \cite{Chinnathambi2018} a regular DNN model is proposed but the model is only evaluated on a test dataset comprising a single day and compared against a simple MLP. In \cite{Jiang2018}, the use of an LSTM model for EPF is evaluated, but the method is only compared with three neural networks and a simple statistical method, and the evaluation is done using only 4 weeks of data. Likewise, \cite{Mujeeb2018} proposes a model based on an LSTM but a comparison against other methods is not done and the test dataset only comprises 2 weeks of data. In \cite{Zhu2018}, another LSTM model is proposed but, as other studies, the test dataset comprises some months of data and the method is only compared against a simple decision tree and a support vector regressor; moreover, the exact split between the training and test dataset is not specified and it is unclear what is exactly the performance of the model. An exception to these studies is \cite{Kolberg2018} which proposes a series of DL models and compares them for a year of data against several advanced statistical methods such as LASSO and a simpler ML method. The main drawbacks of the study are that it is based on a single market and that it only considers a simple ML method as a benchmark. In addition, the study focuses on intraday electricity prices, while most of the literature (including the current paper) considers forecasting day-ahead electricity prices.

In 2019, the main focus of the papers was the same as in 2018: 1) evaluating the performance of different deep recurrent networks (mostly LSTMs) \cite{Zhou2019,Mujeeb2019,Chen2019,Mujeeb2019a,Xu2019,Meier2019,Chang2018}, 2) proposing new hybrid deep learning methods usually based on LSTMs and CNNs \cite{Jahangir2019,Ahmad2019,Xu2019,Aineto2019,Zahid2019,Chang2019,Afrasiabi2019a}, or 3) employing regular DNN models \cite{Atef2019,Luo2019,Schnuerch2019}. Similarly, as with most studies in 2018, the new studies were more limited than \cite{Wang2016b,Lago2018a} as no comparisons with state-of-the-art statistical methods were made and long test datasets were seldom used. In this context, even though some studies \cite{Chen2019,Schnuerch2019} tried to compare the proposed methods with existing DL models \cite{Lago2018a}, they either failed to re-estimated the benchmark models for the new case study \cite{Chen2019} or they overfitted the DL benchmark models \cite{Schnuerch2019}.

 In detail, \cite{Zhou2019} proposes different LSTM models but the new models are only compared against 5 other ML techniques and using a test period of 4 weeks. In \cite{Zahid2019}, a CNN model is proposed but the new model is just compared against three simple ML methods and using a test dataset that comprises a week. In \cite{Mujeeb2019}, a model based on an LSTM is proposed but it is only compared against three simple ML methods and for a period of 12 weeks. In \cite{Atef2019}, the performance of a DNN is compared to that of an SVR model and, as the comparison only includes these two models, it is obviously very limited. In \cite{Luo2019}, a DNN is used as part of a two-step forecasting method; as in many other studies, the comparison is performed for one month of data and limited to two simple ML models (a SVR and an MLP) and a standard linear model. In \cite{Mujeeb2019a}, two DL models are proposed but the models are only compared to very simple ML methods (extreme learning machines and standard MLPs) and using a test dataset spanning eight months. In \cite{Chen2019}, a bidirectional LSTM to forecast prices in the French market is proposed; however, the study only considers historical prices as input features and the proposed method is only compared against DL models and a simple autoregressive model. In addition, the benchmark DL models are copied from \cite{Lago2018a} (a completely different case study that considers exogenous inputs and a different market) without re-tuning the hyperparameters to the new case study. 
 
 In \cite{Schnuerch2019}, a neural network that uses data from order books is proposed and compared against DL methods from the literature, e.g.\ the ones proposed in \cite{Lago2018a}. While the new model outperforms existing DL methods, the DL methods from the literature are trained to overfit the training dataset\footnote{In the training dataset, the proposed model and some naive ML benchmark models yield a root mean square error in the order of 6. For the test dataset, for the same models, that error is between 9 and 12. By contrast, the training error of the benchmark DL model is 2, and the test error is 20. Having a training error that is 1/3 of the error of other models but a test error that is 10 times larger than the training error is a clear sign for overfitting (especially when for the rest of the models the test error is just 1.5 larger than the training error).}. Therefore, the comparison is not meaningful (the DL benchmark models will necessarily perform poorly in the test dataset) and it cannot be assessed how the new model performs. In \cite{Jahangir2019}, a hybrid DL forecasting method is proposed based on stacked denoising autoencoders for pre-training, regular autoencodes for feature selection, and a rough DNN as a forecasting method. As other studies, the method is only compared against other simpler ML models. Moreover, the importance of each of the four modules of the hybrid method is not studied and the study does not re-calibrate the models with new data: the models are trained once and evaluated during a whole year. Similarly, \cite{Ahmad2019} proposes a CNN hybrid model that uses mutual information, random forests, grey correlation analysis, and recursive feature elimination for feature selection. Unlike most models, the algorithm is trained to classify prices instead of predicting their scalar values; however, details of how this process is done are not provided. In addition, the method is only compared against simpler ML methods and evaluated for less than a year of data (the study uses 1 year for testing and training but the split is not specified). Likewise, \cite{Afrasiabi2019a} proposes a hybrid model based on CNNs and RNNs in the context of microgrids; as other studies, the method is evaluated in a small dataset, it is not compared against  state-of-the-art statistical methods, and the exact split between training and test datasets is not specified.

\subsection{Hybrid methods}
Within the field of EPF, the research area that has received most attention in the last 5 years has been hybrid forecasting methods. In this time frame, more than 100 articles proposing new hybrid methods have been published\footnote{This data is based on two searches in Scopus looking for keywords in the title, abstract, and keywords. The first search is based on the following query  \texttt{TITLE-ABS-KEY(((forecast*) OR (predict*)) AND (electricity)  AND  (price*) AND (hybrid))}. The second search is very similar but replacing the keyword \texttt{hybrid} by \texttt{neural AND network}. Note that, while this search is not as complete as the one for DL, it provides enough material for building an overview of the state of the field.}, i.e.\ approximately 5  times more than articles based on DL. Hybrid models are very complex forecasting frameworks that are composed of several algorithms. Usually, they comprise at least two of the following five modules: 
\begin{itemize}
    \item An algorithm for decomposing data.
    \item An algorithm for feature selection.
    \item An algorithm to cluster data.
    \item One or more forecasting models whose predictions are combined.
    \item Some type of heuristic optimization algorithm to either estimate the models or their hyperparameters.
\end{itemize}

In terms of decomposition methods, the most widely used technique is the wavelet transform \cite{Chang2019,Nazar2018,Anamika2018,Gao2018,Bento2018,Peter2016,Singh2017,Yang2017,Olamaee2016,Zhang2015}. Alternatives methods include empirical mode decomposition \cite{Zhang2019,Hong2014a}, variational mode decomposition \cite{AruldossAlbertVictoire2018,Lahmiri2017}, and singular spectrum analysis \cite{Varshney2016,Xiao2017}.

For feature selection, the most commonly used algorithms are correlation analysis \cite{Bento2018,Khajeh2017,Bisoi2018,Kim2015,Hong2014a} and the  mutual information technique \cite{Pourdaryaei2019,Gao2019,Gao2018,Khajeh2017,Ebrahimian2018,Abedinia2015}. Other algorithms include
classification and regression trees with recursive feature elimination \cite{Naz2019} or Relief-F \cite{Naz2019}.

For clustering data, the algorithms are usually based on one of the following four: k-means \cite{Ghayekhloo2019,Itaba2018}, self-organizing maps \cite{Nazar2018,Ghayekhloo2019,Ghofrani2017}, enhanced game theoretic clustering \cite{Ghayekhloo2019}, or fuzzy clustering \cite{Itaba2017,Gao2018}

The selection of forecasting models is much more diverse. The most widely used method is the standard MLP
\cite{Nazar2018,Anamika2018,Bento2018,Khajeh2017,Varshney2016,Xiao2017,Ebrahimian2018,Kim2015,Hong2014a,Zhou2018,Abedinia2015}, followed by the \textit{adaptive network-based fuzzy inference system} (ANFIS) \cite{Zhang2019,Pourdaryaei2019,Nazar2018}, radial basis function network \cite{Itaba2017,Olamaee2016,Zhou2018}, and autoregressive models like ARMA or ARIMA \cite{Zhang2019,Yang2017,Olamaee2016,Zhou2018}. Other models include LSTM \cite{Chang2019}, linear regression \cite{Naz2019}, extreme learning machine \cite{Naz2019,Yang2017}, CNN \cite{Naz2019}, Bayesian neural network \cite{Ghayekhloo2019,Ghofrani2017},   exponential GARCH \cite{Zhang2019}, echo state neural network \cite{AruldossAlbertVictoire2018}, Elman neural networks \cite{Gao2019}, and support vector regressors \cite{Zhou2018}. It is important to note that in many of the approaches, the hybrid method does not consider a single forecasting model but combines several of them \cite{Naz2019,Zhang2019,Zhou2018,Olamaee2016,Nazar2018,Abedinia2015}.

Just as for the forecasting model, the diversity of the heuristic optimization algorithms is also large. While the most often utilized  algorithm  is particle swarm optimization \cite{Pourdaryaei2019,Anamika2018,Itaba2017,Lahmiri2017,Ebrahimian2018,Yang2017}, many other approaches are also used: differential evolution \cite{AruldossAlbertVictoire2018}, genetic algorithm \cite{Pourdaryaei2019}, backtracking search \cite{Pourdaryaei2019}, deterministic annealing \cite{Itaba2017}, bat algorithm \cite{Bento2018}, vaporization precipitation-based water cycle algorithm \cite{Bisoi2018}, cuckoo search \cite{Xiao2017,Kim2015}, or honey bee mating optimization \cite{Olamaee2016}.

In spite of the large number of published works, the research in hybrid methods suffers from the same problems as discussed earlier. First, most of the studies either avoid comparison with well-established methods \cite{Khajeh2017,Singh2018,Lahmiri2017,Peter2016,Singh2017,Darudi2015,Naz2019,Zhang2019,AruldossAlbertVictoire2018,Pourdaryaei2019,Gao2019,Nazar2018,Itaba2017,Zhou2018,Bisoi2018} or resort to comparisons using outdated methodologies \cite{Ghayekhloo2019,Anamika2018,Gao2018,Yang2017,Bento2018,Olamaee2016,Varshney2016,Xiao2017}. Hence, the accuracy of the new proposed methods cannot be accurately established.

Second, the considered studies usually employ very small datasets consisting either of a few days \cite{Chang2019,Gao2019,Nazar2018,Zhou2018,Singh2018,Yang2017} or a few weeks \cite{Olamaee2016,Darudi2015,Yang2017,Ghayekhloo2019,AruldossAlbertVictoire2018,Pourdaryaei2019,Gao2019,Nazar2018,Bisoi2018,Peter2016,Bento2018,Anamika2018,Itaba2017,Khajeh2017,Varshney2016,Xiao2017}. Thus, drawing conclusions is nearly impossible and it is unclear whether the accuracy results are just the outcome of selecting a convenient test period. 

Besides these two problems, for many hybrid methods the effect of selecting variants of the different hybrid components is not analyzed \cite{Olamaee2016,Singh2018,Naz2019,Anamika2018,Gao2018,AruldossAlbertVictoire2018,Zhou2018,Bento2018,Khajeh2017,Varshney2016,Xiao2017,Darudi2015}. Thus, it is not clear how relevant or useful the individual components are. 

\subsection{State-of-the-art models}
\label{sec:accuratemodels}
Because of the described problems when comparing EPF models, it is very hard to establish what are the state-of-the-art methods. Nevertheless, considering the studies performed in the last years, it can be argued that the LEAR is a very accurate (if not the most accurate) linear model. Moreover, it can also be argued that the accuracy of this model can be further improved by transforming the prices using variance stabilizing transformations, combining forecasts obtained for different calibration windows, and/or using long-term seasonal decomposition. 

For the case of ML models, the selection is harder as the existing comparisons are of worse quality. Considering the most complete  benchmark study in terms of forecasting models \cite{Lago2018a}, it seems that a simple DNN with two layers is one of the best ML models. In particular, while more complex models, e.g.\ LSTMs, could potentially be more accurate,  there is at the moment no sound evidence to validate this claim.

In the case of hybrid models, establishing what is the best model is an impossible task. Firstly, while many hybrid methods have been proposed, they have not been compared with each other nor with  the  LEAR or DNN models. Secondly, as most studies do not evaluate the individual influence of each hybrid component, it is also impossible to establish the best algorithms for each hybrid component, e.g.\ it is unclear what are the best clustering, feature selection method, or data decomposition methods.

With that in mind, we will consider the LEAR and the DNN for the proposed open-access benchmark. In particular, not only are these two methods highly accurate, but they are also relatively simple. As such, we think that they are the best benchmarks to compare new complex EPF forecasting methods with.

\section{Open-access benchmark dataset}
\label{sec:ben}
\label{sec:Datasets}
The first contribution of the paper is to provide a large open-access benchmark dataset on which new methods can be tested, together with the day-ahead forecasts of the proposed open-access methods. In this section, we introduce this dataset, which can be accessed\footnote{Note that we do not own the data in the dataset. However, it can be  freely accessed from different websites, e.g.\ the ENTSO-E transparency platform \cite{entsoe}. In this context, the proposed \texttt{python} library \cite{benchmarkwebsite, epftoolboxdoc} provides an interface to easily access the data.} using the \texttt{python} library built for this study.

\subsection{General characteristics}
For a benchmark dataset in EPF to be fair it should satisfy three conditions: 1) comprise several electricity markets so that the capabilities of new models can be tested under different conditions, 2) be long enough so that algorithms can be analyzed  using out-of-sample datasets that span 1-2 years, and 3) be recent enough to include price effects due to the integration of RES.

Based on these conditions, we propose five datasets representing five different day-ahead electricity markets, each of them comprising 6 years of data. The prices of each market have very distinct dynamics, i.e.\ they all have differences in terms of the frequency and existence of negative prices, zeros prices, and price spikes. In addition, as electricity prices depend on exogenous variables, each dataset comprises two additional time series: day-ahead forecasts of two influential exogenous factors that differ from each market. The length of each dataset equals 2184 days, which translates to six years of 364 days or $6\times52=312$ weeks\footnote{Electricity prices have weekly seasonality. Thus, by approximating a year by 52 weeks because we ensure that the metrics are not offset because a certain day, e.g.\ Monday, is harder to predict than the others.}. All available time series are saved using the local time, and the daylight savings are treated by either arithmetically averaging two values from the extra hour or interpolating the neighboring values for the missing observation.

\subsection{Nord Pool}
The first dataset represents the Nord Pool (NP), i.e.\ the European power market of the Nordic countries, and spans from 01.01.2013 to 24.12.2018. The dataset contains hourly observations of day-ahead prices, the day-ahead load forecast, and the day-ahead wind generation forecast. The dataset was constructed using the data freely available on the webpage of the Nordic power exchange Nord Pool \cite{noordpool}. Figure \ref{fig:datasetspjmnp} (top) displays the electricity price time series of the dataset; as can be seen, the prices are always positives, zero prices are rare, and prices spikes seldom occur.

\subsection{PJM}
\begin{figure*}[ht]
	\begin{subfigure}[t]{1\textwidth}
		\begin{center}
			\includegraphics[width=0.9\textwidth]{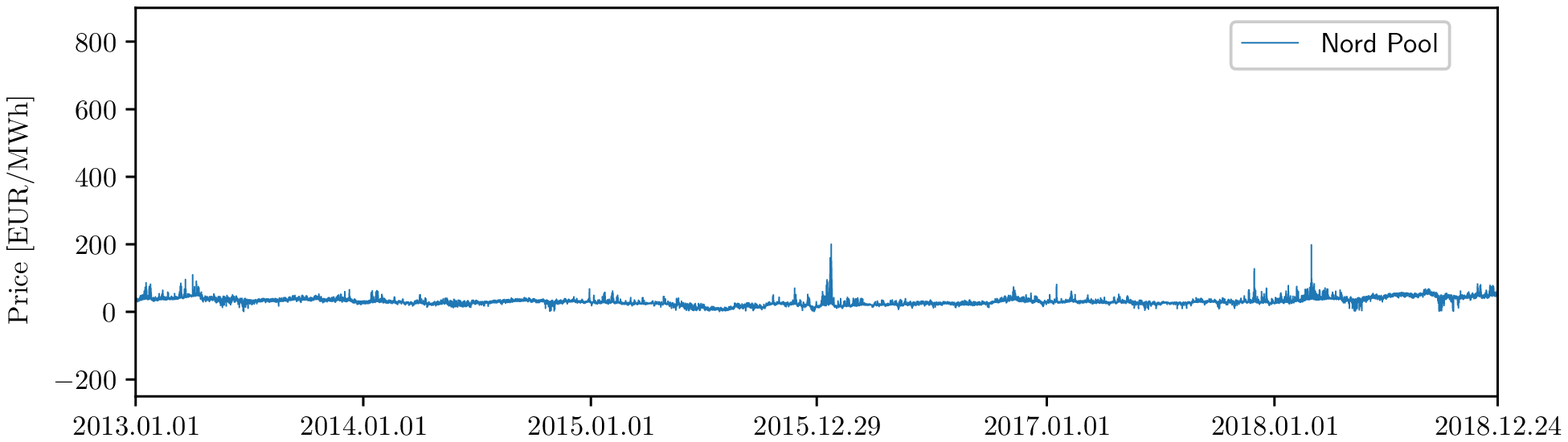}
			\label{fig:data_np}
		\end{center}
	\end{subfigure}\hfill
	\begin{subfigure}[t]{1\textwidth}
		\begin{center}
			\includegraphics[width=0.9\textwidth]{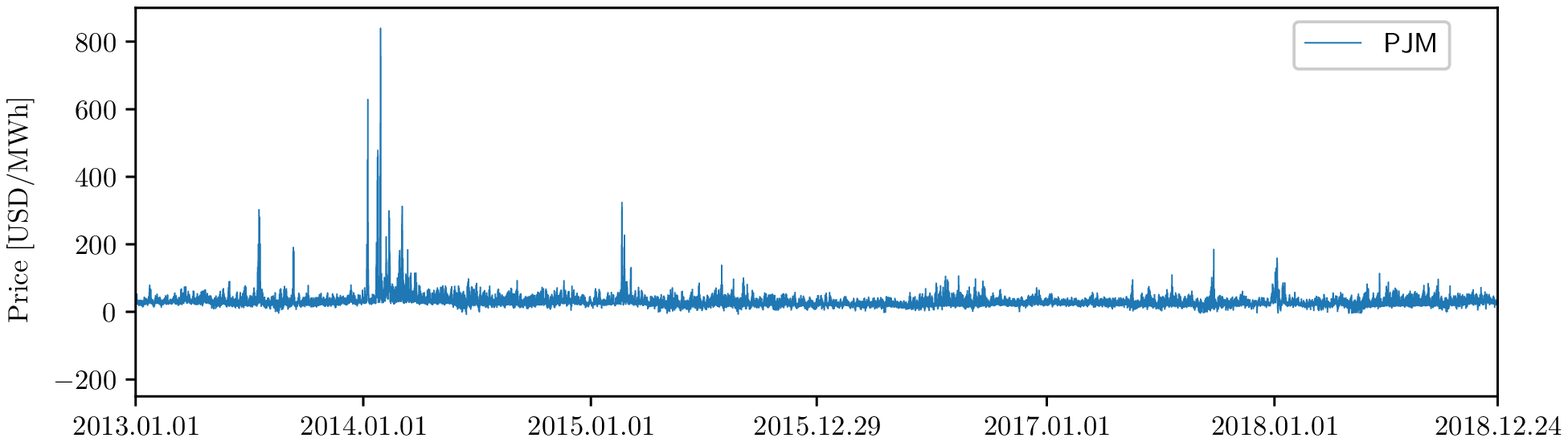}
			\label{fig:data_pjm}			
		\end{center}
	\end{subfigure}
	\caption{Electricity prices time series for two of the five datasets, i.e.\ Nord Pool and PJM, considered in the open-access benchmark dataset (Note that each dataset also includes two exogenous time series that are not plotted here).}
	\label{fig:datasetspjmnp}
\end{figure*}

The second dataset is obtained from the \emph{Pennsylvania-New Jersey-Maryland} (PJM) market in the United States. It covers the same data points as Nord Pool, i.e.\ from 01.01.2013 to 24.12.2018. The three time series are: the zonal prices in the \emph{Commonwealth Edison} (COMED) (a zone located in the state of Illinois) and two day-ahead load forecast series, one describing the system load and the second one the COMED zonal load. The data is freely available on the PJM's website \cite{pjm}. Figure \ref{fig:datasetspjmnp} (bottom) represents the electricity price time series of the dataset; as with the NP market, the prices are always positive and zero prices are rare; however, unlike with the prices in the NP market, spikes appear frequently.

\subsection{EPEX-BE}
The third dataset represents the EPEX-BE market, the day-ahead electricity market in Belgium, which is operated by EPEX SPOT. The dataset spans from 09.01.2011 to 31.12.2016. The two exogenous data series represent the day-ahead load forecast and the day-ahead generation forecast in France. While this selection might be surprising, it has been shown \cite{Lago2018} that these two are the best predictors of Belgian prices. The price data is freely available in the ENTSO-E transparency platform \cite{entsoe} and the ELIA website \cite{Elia}, and the load and generation day-ahead forecasts are freely available in \cite{RTE}. It is important to note that this dataset is particularly interesting because it is harder to predict. Figure \ref{fig:datasetsbefr} (top) shows the electricity price time series of the dataset; unlike the prices in the PJM and NP markets, negative prices and zero prices appear more frequently, and price spikes are very common.

\subsection{EPEX-FR}
The fourth dataset represents the EPEX-FR market, the day-ahead electricity market in France, which is also operated by EPEX SPOT. The dataset spans the same period as the EPEX-BE dataset, i.e.\ from 09.01.2011 to 31.12.2016. Besides the electricity prices, the dataset comprises the day-ahead load forecast and the day-ahead generation forecast. 
As before, the price data is freely obtained from the ENTSO-E transparency platform \cite{entsoe}, and the load and generation day-ahead forecasts are freely available on the webpage of RTE \cite{RTE}, i.e.\ the \textit{transmission system operator} (TSO) in France. Figure \ref{fig:datasetsbefr} (middle) displays the electricity price time series of the dataset; as in the EPEX-BE market, negative prices, zero prices, and spikes are very common.

\subsection{EPEX-DE}
\begin{figure*}[!ht]
	\begin{subfigure}[t]{\textwidth}
		\begin{center}
			\includegraphics[width=0.9\textwidth]{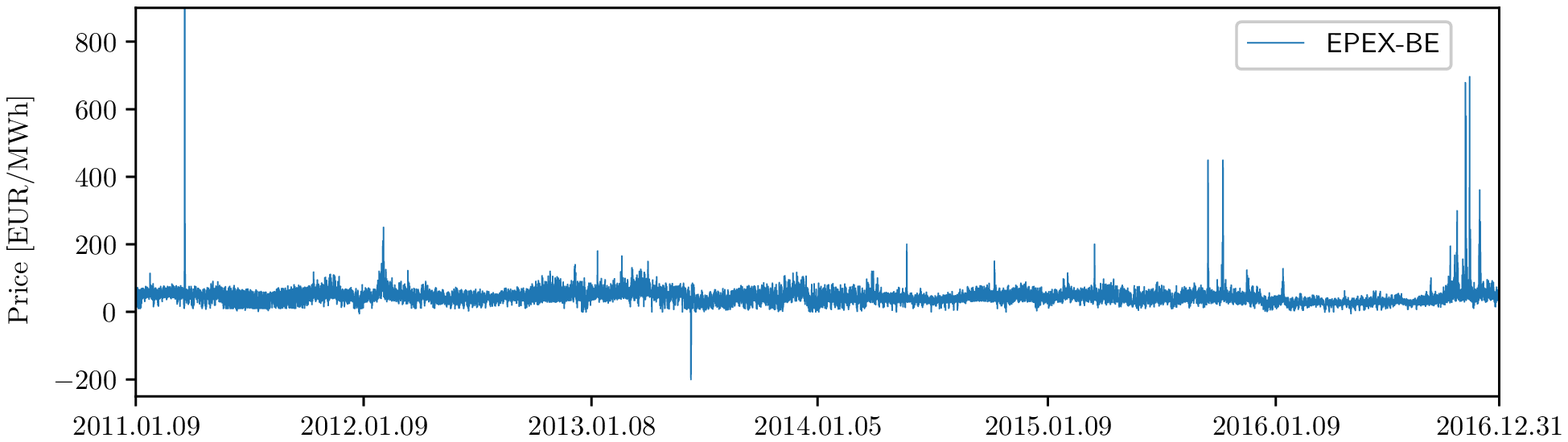}
			\label{fig:data_be}
		\end{center}
	\end{subfigure}\hfill
	\begin{subfigure}[t]{\textwidth}
		\begin{center}
			\includegraphics[width=0.9\textwidth]{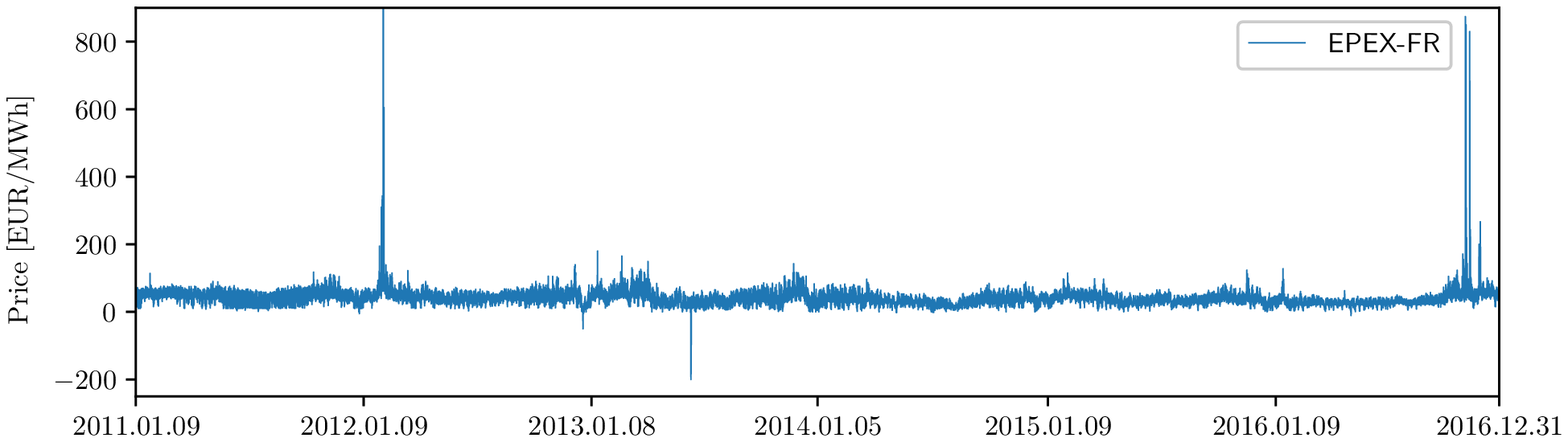}
			\label{fig:data_fr}	
		\end{center}
	\end{subfigure}
	\begin{subfigure}[t]{\textwidth}
		\begin{center}
			\includegraphics[width=0.9\textwidth]{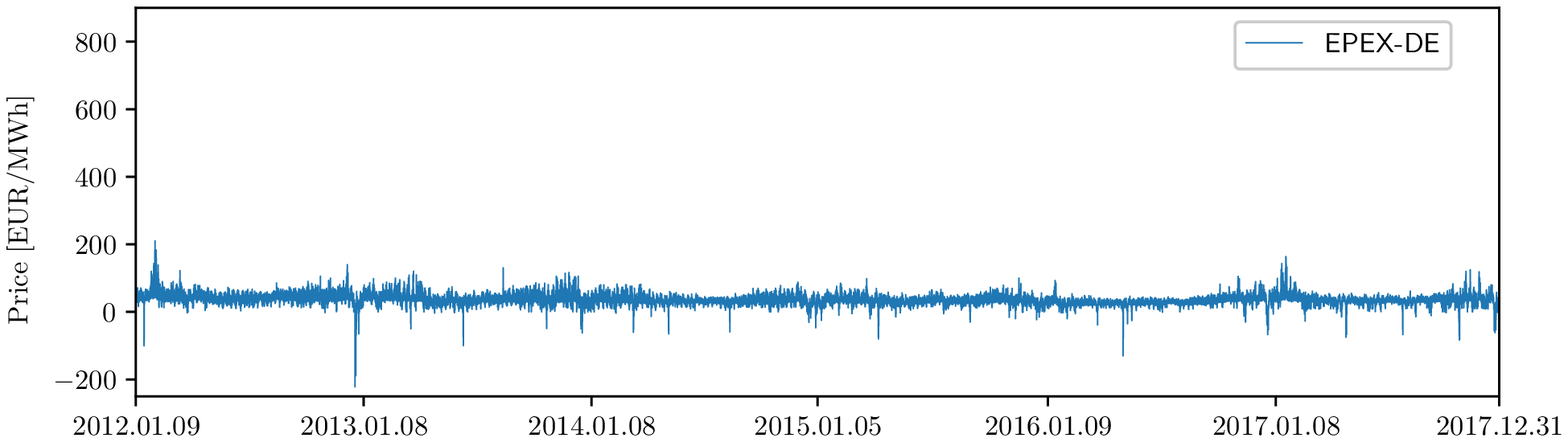}
			\label{fig:data_de}			
		\end{center}
	\end{subfigure}
	\caption{Electricity prices time series for three of the five datasets, i.e.\ EPEX-BE, EPEX-FR, and EPEX-DE, considered in the open-access benchmark dataset (Note that each dataset also includes two exogenous time series that are not plotted here). The EPEX-BE and EPEX-FR time series are similar because the EPEX-FR and EPEX-BE are highly coupled markets \cite{Lago2018}. To keep the plots readable, the upper limit of the y-axis is below the maximum price; this only affects one spike in EPEX-FR and another one in EPEX-BE.}
	\label{fig:datasetsbefr}
\end{figure*}

The last dataset describes the EPEX-DE market, the German electricity market, which is also operated by EPEX SPOT. The dataset spans from 09.01.2012 to 31.12.2017. Besides the prices, the dataset comprises the day-ahead zonal load forecast in the TSO Amprion zone and the aggregated day-ahead wind and solar generation forecasts in the zones of the 3 largest\footnote{There are 4 TSOs in Germany.} TSOs (Amprion, TenneT, and 50Hertz). The price data is freely obtained from the ENTSO-E transparency platform \cite{entsoe}, the zonal load day-ahead forecasts is freely available in the website of Amprion \cite{amprion}, and the wind and solar forecasts in the websites of Amprion \cite{amprion}, 50Hertz \cite{50hertz}, and TenneT \cite{tennet}. Figure \ref{fig:datasetsbefr} (bottom) displays the electricity price time series of the dataset; as can be seen, while negative and zero prices occur more often than in the other four markets, price spikes are more rare.

\subsection{Training and testing periods}
\label{sec:traingtest}
For each dataset, the testing period is defined as the last 104 weeks, i.e.\ the last two years, of the dataset. The  exact dates of the testing datasets are defined in Table \ref{tab:datestesting}.

\begin{table}[htb]
	\centering
	\def\arraystretch{1.3}%
	\caption{Start and end dates of the testing (out-of-sample) datasets for each electricity market.}
	\label{tab:datestesting}
	\begin{tabular}{l|c}
	\textbf{Market}	& \textbf{Test period}\\
		\hline
		{Nord pool}	& 27.12.2016 -- 24.12.2018 \\
		{PJM}	& 27.12.2016 -- 24.12.2018 \\
		{EPEX-FR}	&  04.01.2015 -- 31.12.2016 \\
		{EPEX-BE}&  04.01.2015 -- 31.12.2016 \\
		{EPEX-DE}&  04.01.2016 -- 31.12.2017
	\end{tabular}
\end{table}

\noindent It is important to note that, as we will argue in Section \ref{sec:guidelines}, selecting two years as the testing period is paramount to ensure good research practices in EPF.

Unlike the testing dataset, the training dataset cannot be defined as it will vary between different models. In, general, the training dataset will comprise any data that is prior to the data under study. However, the exact data will change depending on two concepts, i.e.\ calibration window and recalibration:

\begin{itemize}
	\item While there are four years of data available for estimating the model, it might be desirable to employ only recent data, e.g.\ to avoid estimating effects that no longer play a role. The amount of past data employed for estimation defines the calibration window.
	\item The model can be estimated once and then evaluated for the full test dataset, or it can be continuously recalibrated on daily basis to incorporate the input of recent data.
\end{itemize}

\noindent For example, let us consider predicting the prices in the NP on 15.02.2017. A model using a calibration window of 52 weeks and no recalibration would employ a training dataset comprising the data between 29.12.2016 and 26.12.2016, i.e.\ one year prior to the start of the test period. By contrast, a model using a calibration window of 104 weeks and daily recalibration would employ the data between 18.02.2015 and 14.02.2017.

\section{Open-access benchmark models}
\label{sec:mod}
\label{sec:Models}
The second contribution of the paper is to provide a set of state-of-the-art open-source forecasting methods as an open-source \texttt{python} toolbox. As explained in Section \ref{sec:accuratemodels}, the   LEAR \cite{Uniejewski2016} and the DNN \cite{Lago2018a} models are not only highly accurate but also relatively simple. Therefore, we implement these two methods and provide their code freely available as part of the proposed toolbox \cite{benchmarkwebsite,epftoolboxdoc}. It is important to note that the use of the proposed open-access methods is fully documented and automated so researchers can test and use them without  expert knowledge. 

For the sake of simplicity, the description provided here is limited to the bare minimum. For further details on the two models we refer to the original papers \cite{Uniejewski2016,Lago2018a}.

\subsection{Input features}
\label{sec:input_features_fARX_DNN}
Before describing each model, let us define the input features that are considered. Independently of the model, the available input features to forecast the 24 day-ahead prices of day $d$, i.e.\ $\vc{p}_{d} = [p_{d,1},\ldots,p_{d,24}]^\top$, are the same:

\begin{itemize}
	\item Historical day-ahead prices of the previous three days and one week ago, i.e.\ $\vc{p}_{d-1}$, $\vc{p}_{d-2}$, $\vc{p}_{d-3}$, $\vc{p}_{d-7}$.
	\item The  day-ahead forecasts of the two variables of interest (see Section \ref{sec:Datasets} for details) for day $d$ available on day $d-1$, i.e.\ $\vc{{x}}^1_d=[x^1_{d,1},\ldots,x^1_{d,24}]^\top$ and $\vc{{x}}^2_d=[x^2_{d,1},\ldots,x^2_{d,24}]^\top$; note that the variables of interest are different for each market.
	\item Historical day-ahead forecasts of the variables of interest 
	the previous day and one week ago, i.e.\ $\vc{{x}}^1_{d-1}$, $\vc{{x}}^1_{d-7}$, $\vc{{x}}^2_{d-1}$, $\vc{{x}}^2_{d-7}$.
	\item A dummy variable $\vc{z}_d$ that represents the day of the week. In the case of the linear model, following the standard practice in the literature \cite{Uniejewski2018,Uniejewski2016,Marcjasz2018}, this is a binary vector $\vc{z}=[z_{d,1}, \ldots, z_{d,7}]^\top$ that encodes every day of the week by setting all elements to zero except the element that identifies the day of the week, e.g.\ $[1, 0, 0, 0, 0, 0, 0]$ represents Monday and $[0, 1, 0, 0, 0, 0, 0]$ Tuesday. In the case of the neural network, for the sake of simplicity, the day of the week is modeled with a multi-value input $z_d\in\{1,\ldots,7\}$. 
\end{itemize}

In total, we consider a total of 247 available input features for each LEAR model and 241 input features for each DNN model. It is important to note that, while the available input features are the same, each method performs a different feature selection procedure:
\begin{itemize}
    \item Each of the LEAR models finds the optimal set of features using LASSO as an embedded feature selection, i.e.\ each of the models uses L1-regularization to select among the 247 features.
    \item For the DNN, as in the original study \cite{Lago2018a}, the input features are optimized together with the hyperparameters using the tree Parzen estimator \cite{Bergstra2011} (see Section \ref{sec:dnn_model_explanation} for details).
\end{itemize}

\noindent In both cases, the feature selection is fully automated and does not require expert intervention.

\subsection{The LEAR model}
\label{sec:farxmodel}
The first model is the LEAR model \cite{Uniejewski2016}, a parameter-rich ARX model estimated using LASSO  as an implicit feature selection approach. To enhance the model as shown by \cite{Uniejewski2018a}, the data is preprocessed with the \textit{arc hyperbolic sine} (asinh) variance stabilizing transformation. Long-term seasonal decomposition is not considered for the sake of simplicity; particularly, while it has been shown to further improve the performance of the LEAR, we leave it out for future research.

As in \cite{Marcjasz2018}, to further enhance the model, we recalibrated daily over different calibration window lengths: 8 weeks, 12 weeks, 3 years, and 4 years. We consider short windows (8-12 weeks) in combination with long windows (3-4 years) because it has been empirically shown to lead to better results \cite{Marcjasz2018}. In this context, we consider a minimum of 8 weeks as lower windows might not have enough information to correctly estimate parameter-rich models \cite{Marcjasz2018}.

The LEAR model to predict price $p_{d,h}$ on day $d$ and hour $h$ is defined by: 
\begin{alignat}{4}
{p}_{d,h} = &f(\vc{p}_{d-1},\vc{p}_{d-2}, \vc{p}_{d-3}, \vc{p}_{d-7},\vc{x}^{i}_{d},\vc{x}^{i}_{d-1},\vc{x}^{i}_{d-7},\bm{\theta}_{h}) + \varepsilon_{d,h}\span\span\span\nonumber\\
=&\sum_{i=1}^{24}\theta_{h,i}\cdot{p}_{d-1,i} \,&+&\, \sum_{i=1}^{24}\theta_{h,24+i}\cdot{p}_{d-2,i} \nonumber\\ 
&+\sum_{i=1}^{24}\theta_{h,48+i}\cdot{p}_{d-3,i} \,&+&\, \sum_{i=1}^{24}\theta_{h,72+i}\cdot{p}_{d-7,i}\nonumber\\
&+\sum_{i=1}^{24}\theta_{h,96+i}\cdot{x}^1_{d,i} \,&+&\, \sum_{i=1}^{24}\theta_{h,120+i}\cdot{x}^2_{d,i}\nonumber\\
&+\sum_{i=1}^{24}\theta_{h,144+i}\cdot{x}^1_{d-1,i} \,&+&\, \sum_{i=1}^{24}\theta_{h,168+i}\cdot{x}^2_{d-1,i}\nonumber\\ 
&+\sum_{i=1}^{24}\theta_{h,192+i}\cdot{x}^1_{d-7,i} \,&+&\, \sum_{i=1}^{24}\theta_{h,216+i}\cdot{x}^2_{d-7,i}\nonumber\\  
&+ \sum_{i=1}^{7}\theta_{h,240+i}\cdot z_{d,i} \,&+&\, \varepsilon_{d,h}\label{eqn:fARX}
\end{alignat}

\noindent where $\bm{\theta}_{h} = [\theta_{h,1},\ldots,\theta_{h,247}]^\top$ are the 247 parameters of the LEAR model for hour $h$. Many of these parameters become zero when \eqref{eqn:fARX} is estimated using LASSO:
\begin{align}
\hat{\boldsymbol\theta}_h = \,&
\underset{\boldsymbol\theta_h}{\textrm{argmin}} \left\{\text{{RSS}} + \lambda\,\left\|\boldsymbol\theta_h\right\|_1\right\} \nonumber \\
= \,& \underset{\boldsymbol\theta_h}{\textrm{argmin}} \left\{\text{{RSS}} + \lambda\sum_{i=1}^{247} \left|\theta_{h,i}\right|\right\}, \label{eqn:LASSO}
\end{align}
where $\text{RSS}=\sum_{d=8}^{N_d}(p_{d,h}-\hat{p}_{d,h})^2$ is the  sum of squares residuals, $\hat{p}_{d,h}$ the price forecast, $N_d$ is the number of days in the training dataset, and $\lambda\ge0$ is the \textit{tuning} (or \textit{regularization}) hyperparameter of LASSO.
Due to the computational speed of estimating with LASSO, during every daily recalibration, the hyperparameter $\lambda$ that regulates the L$_1$ penalty is optimized. This can be done using an \emph{ex-ante} cross-validation procedure \cite{Hastie2001}. In this study, to further reduce the computational cost, we propose an efficient hybrid approach to perform the optimal selection of $\lambda$. See Section \ref{sec:lambdaselection} for details.

\subsubsection{Regularization hyperparameter}
The hyperparameter $\lambda$ of LASSO can be optimized in multiple ways, each one of them with different merits and disadvantages. A first approach is to optimize $\lambda$ once and then keep it fixed for the whole test period. Although it requires very low computation costs, the limitation of this  approach is that it assumes that the hyperparameter $\lambda$ does not change over time. This assumption might hinder the performance of the estimator as the regularization parameter does not change even when the market might do. 

A second approach is to recalibrate the hyperparameter on a periodic basis using a validation dataset. Although this method yields good results, tuning  the recalibration frequency and calibration window is complicated, the computational cost is large, and the results may vary between datasets \cite{Uniejewski2018}. 

A third option is to recalibrate the hyperparameter periodically, but using \textit{cross-validation} (CV): splitting the data into disjoint partitions, using each possible partition once as a test dataset with the remaining data as the training dataset, and selecting the hyperparameter that performs the best across all partitions \cite{Hastie2001}. Although this approach is highly accurate, its computation costs are very large.

A fourth option is to periodically update the hyperparameter but using information criteria, e.g.\ the \textit{Akaike information criterion} (AIC) or the Bayesian information criterion \cite{Ziel2018,Ziel2015a,Ziel2016}. As before, this involves training multiple LASSO models to compute the information criteria for each possible hyperparameter value, which in turn leads to a high computational cost.

Lastly, one can use the \textit{least angle regression} (LARS) LASSO \cite{Efron2004} for estimating the model instead of the coordinate descent implementation. This estimation procedure has the advantage of computing the whole LASSO solution path, which in turn allows to compute the information criteria or perform CV much faster. 

\subsubsection{Selecting the regularization hyperparameter}
\label{sec:lambdaselection}
To select $\lambda$ we propose a hybrid approach. On a daily basis, we estimate the hyperparameter using the LARS method with the in-sample AIC. Then, using the optimal $\lambda$ obtained from the LARS method, we recalibrate the LEAR using the traditional coordinate descent implementation. 


The reason for proposing this hybrid approach is that it provides a good trade-off between computational complexity and accuracy. In particular, it leverages the computational efficiency of LARS for ex-ante $\lambda$ selection with the predictive performance on short calibration windows of the coordinate descent LASSO. 

It is important to note that we have studied multiple approaches to select $\lambda$: (i) daily recalibration, CV, with coordinate descent; (ii) daily recalibration, CV, with LARS;  (iii) daily recalibration with LARS and AIC. However, the computational cost of the first method was too high (in the same order of magnitude as the cost of the DNN model), and the accuracy of the other two was not good. By contrast, the proposed approach had a performance on par with coordinate descent LASSO using CV, but with a computational cost that was an order of magnitude lower.

\subsection{The DNN model}
\label{sec:dnn_model_explanation}
The second model is the DNN \cite{Lago2018a}, one of the most simple DL models whose input features and hyperparameters can be optimized and tailored for each case study without the need of expert knowledge.

\subsubsection{Structure}
\begin{figure}[htb]
	\begin{center}
		\def\Nin{2}
\def\Nhid{3}
\def\Nhidd{3}
\def\Nout{2}
\setlength{\separ}{2cm}
\begin{tikzpicture}[shorten >=1pt,->,draw=black!50]

\tikzstyle{every pin edge}=[<-,shorten <=1pt]
\tikzstyle{neuron}=[circle,fill=black!25,minimum size=17pt,inner sep=0pt]
\tikzstyle{input neuron}=[neuron, fill=green!40];
\tikzstyle{output neuron}=[neuron, fill=red!50];
\tikzstyle{hidden neuron}=[neuron, fill=blue!50];
\tikzstyle{hidden neuron 2}=[neuron, fill=orange!50];
\tikzstyle{annot} = [text width=4em, text centered]

\foreach \name / \y in {1,...,\Nin}
\node[input neuron] (I-\name) at (0,-\y) {$p_{d-1,\y}$};
\node at (0,-\Nin-1) {$\vdots$};
\node[input neuron] (I-\Nin+1) at (0,-\Nin-2) {$x^2_{d,24}$};

\foreach \name / \y in {1,...,\Nhid}
\path[yshift=0.5cm]
node[hidden neuron] (H-\name) at (\separ,-\y cm) {$\hid_{1\y}$};
\node at (\separ,-\Nhid-0.25) {$\vdots$};
\node[hidden neuron] (H-\Nhid+1) at (\separ,-\Nhid-1.25) {$\hid_{1\nhidn_1}$};

\foreach \name / \y in {1,...,\Nhidd}
\path[yshift=0.5cm]
node[hidden neuron 2] (H2-\name) at (2\separ,-\y cm) {$\hid_{2\y}$};
\node at (2\separ,-\Nhidd-0.25) {$\vdots$};
\node[hidden neuron 2] (H2-\Nhidd+1) at (2\separ,-\Nhidd-1.25) {$\hid_{2\nhidn_2}$};

\foreach \name / \y in {1,...,\Nout}
\path[yshift=0cm]
node[output neuron] (O-\name) at (3\separ,-\y cm) {$p_{\y}$};
\node at (3\separ,-\Nout-1) {$\vdots$};
\node[output neuron] (O-\Nout+1) at (3\separ,-\Nout-2) {$p_{{24}}$};



\foreach \source in {1,...,\Nin}
\foreach \dest in {1,...,\Nhid}
\path (I-\source) edge (H-\dest);

\foreach \dest in {1,...,\Nhid}
\path (I-\Nin+1) edge (H-\dest);
\foreach \source in {1,...,\Nin}
\path (I-\source) edge (H-\Nhid+1);
\path (I-\Nin+1) edge (H-\Nhid+1);

\foreach \source in {1,...,\Nhid}
\foreach \dest in {1,...,\Nhidd}
\path (H-\source) edge (H2-\dest);

\foreach \dest in {1,...,\Nhidd}
\path (H-\Nhid+1) edge (H2-\dest);
\foreach \source in {1,...,\Nhid}
\path (H-\source) edge (H2-\Nhidd+1);
\path (H-\Nhid+1) edge (H2-\Nhidd+1);

\foreach \source in {1,...,\Nhidd}
\foreach \dest in {1,...,\Nout}
\path (H2-\source) edge (O-\dest);

\foreach \dest in {1,...,\Nout}
\path (H2-\Nhidd+1) edge (O-\dest);
\foreach \source in {1,...,\Nhid}
\path (H2-\source) edge (O-\Nout+1);
\path (H2-\Nhidd+1) edge (O-\Nout+1);

\node[annot,above of=H-1, node distance=1cm] (hl) {Hidden layer};
\node[annot,above of=H2-1, node distance=1cm] (hl) {Hidden layer};
\node[annot,above of=I-1, node distance=1cm] {Input layer};
\node[annot,above of=O-1, node distance=1cm] {Output layer};

\node at (3\separ,-\Nout-1) {$\vdots$};
\end{tikzpicture}
		\caption{Visualization of a sample DNN model}
		\label{fig:DNN}
	\end{center}
\end{figure}
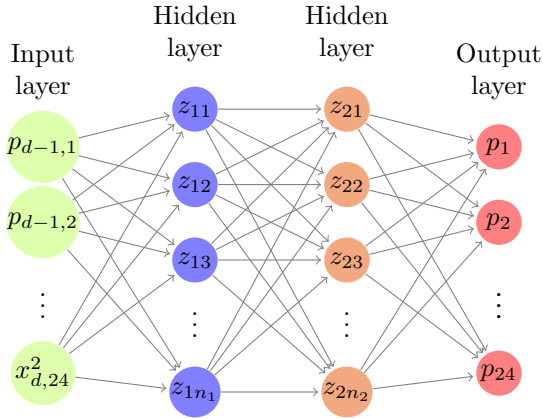

The DNN is a deep feedforward neural network that contains 4 layers, employs the multivariate framework (single model with 24 outputs), is estimated using Adam, and its hyperparameters and input features are optimized using the tree Parzen estimator \cite{Bergstra2011}, i.e.\ a Bayesian optimization algorithm. The DNN model is visualized in Figure \ref{fig:DNN}.

\setlength{\figW}{\textwidth}
\setlength{\figH}{0.5\figW}

\subsubsection{Training dataset}
For estimating the hyperparameters, the training dataset is fixed and comprises the four years prior to the testing period. For evaluating the testing dataset, the DNN is recalibrated on a daily basis using a calibration window of four years. 

In all cases, the training dataset is split into a training and a validation dataset, with the latter being used for two purposes: performing early stopping \cite{Yao2007} to avoid overfitting and optimizing hyperparameters/features. While the validation dataset alwawys comprises 42 weeks, the split between the training and validation datasets depends on whether the validation dataset is used for hyperparameter/feature selection or for the recalibration step:

\begin{itemize}
    \item For estimating the hyperparameters, as the validation dataset is used to guide the optimization process, the validation dataset is selected as the last 42 weeks of the training dataset. This is done to keep the training and validation datasets completely independent and to avoid overfitting\footnote{Similar as it is done when splitting the dataset between the training and test dataset.}. 
    \item For the testing phase, as the validation dataset is only used for early stopping, it is defined by randomly selecting 42 weeks out of the total 208 weeks employed for training. This is done to ensure that the dataset used for optimizing the DNN parameters includes up-to-date data\footnote{For hyperparameter optimization, as the validation dataset represents the most recent weeks of data, the neural network is trained with data that is almost one year old. While this is not a big problem when decididing on the DNN structure, it should be avoided during  testing to ensure that the DNN captures new market effects.}.
\end{itemize}

As example, let us consider the training and evaluation of a DNN in the Nord Pool market. Before evaluating the DNN, the hyperparameter and features of the DNN are optimised. For that, the employed dataset  comprises the data between 01.01.2013 and 26.12.2016, of which the training dataset represents the first 166 weeks, i.e.\ 01.01.2013 to 07.03.2016, and the validation dataset the last 42, i.e.\ 08.03.2016 to 26.12.2016. During the evaluation of the model, i.e.\ after the hyperparameter and feature selection, the  training and validation datasets comprise the last four years of data but are randomly suffled. For example, to evaluate the DNN during 15.02.2017, the training and validation datasets would represent the data between 20.02.2013 and 14.02.2017, of which 166 randomly selected weeks would define the training dataset and the remaining 42 the validation dataset.

\subsubsection{Hyperparameter and feature selection}
\label{sec:hyperoptDNN}
As in the original DNN paper \cite{Lago2018a}, the hyperparameters and input features are optimized together using the tree-structured Parzen estimator \cite{Bergstra2011}, a Bayesian optimization algorithm based on sequential model-based optimization. To do so, the features are modeled as hyperparameters, with each hyperparameter representing a binary variable that selects whether or not a specific feature is included in the model (as explained in \cite{Lago2018}). In more detail, to select which of the 241 available input features are relevant, the method employs 11 decision variables, i.e.\ 11 hyperparameters:
\begin{itemize}
	\item Four binary hyperparameters (1-4) that indicate whether or not to include the historical day ahead prices $\vc{p}_{d-1}$, $\vc{p}_{d-2}$, $\vc{p}_{d-3}$, $\vc{p}_{d-7}$.  The selection is done per day\footnote{This is done for the sake of simplicity to speed up the optimization procedure of the feature selection. In particular, an alternative could be to use a binary hyperparameter for each individual historical prices; however, is most markets, that would mean using 24 as many hyperparameters as there are 24 different prices per day.}, e.g.\ the algorithm either selects all the prices $\vc{p}_{d-j}$ of $j$ days ago or it cannot select any price from day $d-j$, hence the four hyperparameters.
	\item Two binary hyperparameters (5-6) that indicate  whether or not to include each of the day-ahead forecasts $\vc{x}^1_{d}$ and $\vc{x}^2_{d}$. As with the past prices, this is done for the whole day, i.e.\ a hyperparameter either selects all the elements in $\vc{x}^j_{d}$ or none.
	\item Four binary hyperparameters (7-10) that indicate whether or not to include the historical day-ahead forecasts $\vc{x}^1_{d-1}$, $\vc{x}^2_{d-1}$, $\vc{x}^1_{d-7}$, and $\vc{x}^2_{d-7}$. This selection is also done per day.
	\item One binary hyperparameter (11) that indicates whether or not to include the variable $z_d$ representing the day of the week.
\end{itemize}

\noindent In short, 10 binary hyperparameters indicating whether or not to include 24 inputs each and another binary hyperparameter indicating whether or not to include a dummy variable.

Besides selecting the features, the algorithm also optimizes eight additional hyperparameters: 1) the number of neurons per layer, 2) the activation function, 3) the dropout rate, 4) the learning rate, 5) whether or not to use batch normalization, 6) the type of data preprocessing technique, 7) the initialization of the DNN weights, and 8) the coefficient for L1 regularization that is applied to each layer's kernel.

Unlike the weights of the DNN that are recalibrated on a daily basis, the hyperparameter and features are optimized only once using the four years of data prior to the testing period. It is important to note that the algorithm runs for a number $T$ of iterations, where at every iteration the algorithm infers a potential optimal subset of hyperparameters/features and evaluates this subset in the validation dataset. For the proposed open-access benchmark models, $T$ is selected as 1500 iterations to obtain a trade-off between accuracy and computational requirements\footnote{It can be empirically observed that the performance of the models barely improves after 1000 iterations. Moreover, performing 1500 iteration takes approximately just one day on a regular quadcore laptop like the i7-6920HQ, a computation cost very acceptable when the algorithm has to run only once.}.

\subsection{Ensembles}
\label{sec:ensemblesintro}
For the open-access benchmark, in order to have benchmark predictions  when evaluating ensemble techniques, we also proposose ensembles of LEAR and DNNs as open-access benchmarks of ensembles methods. For the LEAR, the ensemble is built as the arithmetic average of forecasts across four calibration window lengths: 8 weeks, 12 weeks, 3 years, and 4 years. For the DNN, the ensemble is built as the arithmetic average of four different DNNs that are estimated by running the hyperparameter/feature selection procedure four times. In particular, the hyperparameter optimization is asymptotically deterministic, i.e.\ the global optimum is found for an infinite number of iterations. However, for a finite number of iterations and using a different initial random seed, the algorithm is non-deterministic and every run provides a different set of hyperparameters and features. Although each of these hyperparameter/feature subsets represent a local minimum, it is impossible to establish which of the subsets is better as their relative performance on the validation dataset is nearly identical. This effect can be explained due to the DNN being a very flexible model and thus different network architectures being able to obtain equally good results.

\subsection{Software implementation}
The proposed open-access models are developed in \texttt{python}: the LEAR is implemented using the \texttt{scikit-learn} library \cite{scikit-learn} and the DNN model using the \texttt{Keras} library \cite{Chollet2015}. The reason for selecting \texttt{python} is that it is one of the most widely used programming languages, especially in the context of ML and statistical inference. 


\section{Guidelines and best practices in EPF}
\label{sec:guidelines}
As motivated in the introduction, the field of EPF suffers from several problems that prevent having reproducible research and establishing strong conclusions on what methods work best. In this section, we outline some of these issues  and provide some guidelines on how to address them.

\subsection{Length of the test period}
A common practice in EPF is to evaluate new methods on very short test periods. The typical approach is to evaluate the method on 4 weeks of data \cite{Ghayekhloo2019,Pourdaryaei2019,Gao2019,Nazar2018,Anamika2018,Bento2018,Khajeh2017,Bisoi2018,Ghofrani2017,Peter2016,Varshney2016,Xiao2017,Ebrahimian2018,Yang2017,Olamaee2016,Darudi2015,Kim2015,Zhou2019,Jiang2018,Aineto2019}, with each week representing one of the four seasons in the year. This is problematic for three reasons:

\begin{itemize}
	\item Selecting four weeks can lead to cherry-picking the weeks where a given method excels, e.g.\ a method that performs bad with spikes could be evaluated in a week with fewer spikes, leading in turn to biased estimations of the forecasting accuracy. While this is an ethical issue that most researchers would avoid, establishing four weeks testing periods as the standard does  facilitate the malpractice and it should be avoided.
	\item Assuming that the four weeks are randomly selected and no bias is introduced in the selection, it is still not possible to guarantee that these four weeks are representative of the price behavior on a whole year. Particularly, even within a given season, the price dynamics can change dramatically, e.g.\ during winter there are weeks with a lot of sun and wind but there are also weeks without them. Therefore, picking only a week per season rarely represents the average performance of a forecaster in a give dataset.
	\item There are situations in the electrical grid that do not occur very often but that can have a very large effect on electricity prices, e.g.\ when several power plants are under maintenance at the same time. Forecasting methods need to be evaluated under those conditions to ensure that they are also accurate under extreme events. By selecting four weeks most of these effects are neglected.
\end{itemize}

To avoid this problem, we recommend using a minimum of one year as a testing period. This ensures that forecasting methods are evaluated considering the complete set of effects that take place during the year. To guarantee that all researchers have access to this type of data, the open-access benchmark dataset that we propose contains data from several markets and employs a testing period of two years. In addition, the open-access benchmark can be directly accessed using the proposed \texttt{epftoolbox} library \cite{benchmarkwebsite,epftoolboxdoc}.

\subsection{Benchmark models}
A second issue with many EPF publications is that new methods are not compared with well-established methods \cite{Khajeh2017,Singh2018,Lahmiri2017,Peter2016,Singh2017,Darudi2015,Naz2019,Zhang2019,AruldossAlbertVictoire2018,Pourdaryaei2019,Gao2019,Nazar2018,Itaba2017,Zhou2018,Bisoi2018,Chen2019,Zhang2018,Chinnathambi2018,Atef2019,Mujeeb2018,Afrasiabi2019a} or resort to comparisons using either outdated methodologies or simplified methods \cite{Ghayekhloo2019,Anamika2018,Gao2018,Yang2017,Bento2018,Olamaee2016,Varshney2016,Xiao2017,Kuo2018,Ugurlu2018,Jiang2018,Zhou2019,Zahid2019,Mujeeb2019,Luo2019,Mujeeb2019a,Jahangir2019,Ahmad2019,Zhu2018}.

This poses a problem since it becomes very hard to establish which algorithms work best and which ones do not. To address this issue, we recommend using well-established state-of-the-art open-source methods and a common benchmark dataset. With that in mind,
we have provided and make freely available an open-access benchmark dataset comprising 5 markets (as described in Section \ref{sec:Datasets}), and we have implemented, thoroughly tested, and made freely available two state-of-the-art forecasting methods (as described in Section \ref{sec:Models}) and their day-ahead predictions for all 5 datasets over a period of two years (as described in Section \ref{sec:eval}). Additionally, we have implemented all these resources in an easy-to-use toolbox \cite{benchmarkwebsite} and built an adequate documentation \cite{epftoolboxdoc}.


\subsection{Open-access}
A third issue in the field of EPF is that datasets are usually not made publicly available and the code of the proposed methods is not shared. This poses four obvious problems:

\begin{itemize}
	\item Research cannot be reproduced as data is not available. This goes against one of the main principles of science as all research should be reproducible. 
	\item The progress of EPF research is hindered since it is hard to establish which methodologies work well. Consequently, researchers spend unnecessary time re-evaluating methodologies that have been evaluated already.
	\item Comparing new methods with published ones becomes very challenging because researchers have to re-implement methods from the literature. As a result, comparisons with state-of-the-art methods are often avoided, and new methods are usually compared with simple and easy-to-implement methods.
	\item When new methods are proposed, they cannot be compared with published methods under the same circumstances. This leads to comparisons under different conditions and opens up the door to wrong implementations of the original methods, which in turn leads to results that are not correct.
\end{itemize}

As these problems are critical, we directly try to address them by providing an open-access benchmark/toolbox comprising five datasets, two state-of-the-art methods, and a set of day-ahead forecasts of the latter two methods. In addition, we encourage researchers in EPF to share the developed codes and to either share their datasets or use an open-access benchmark dataset.

\subsection{Evaluation metrics for point forecasts}
\label{sec:metricsguidelines}
In the field of EPF, the most widely used metrics to measure the accuracy of point forecasts are the \textit{mean absolute error} (MAE), the \textit{root mean square error} (RMSE), and the \textit{mean absolute percentage error} (MAPE):

\begin{align}
	\mathrm{MAE} &= \frac{1}{24\,N_\mathrm{d}}\sum_{d=1}^{N_\mathrm{d}}\sum_{h=1}^{24}|p_{d,h}-\hat{p}_{d,h}|,\\
	\mathrm{RMSE} &= \sqrt{\frac{1}{24\,N_\mathrm{d}}\sum_{d=1}^{N_\mathrm{d}}{\sum_{h=1}^{24} (p_{d,h}-\hat{p}_{d,h})^2}},\\
	\mathrm{MAPE} &= \frac{1}{24\,N_\mathrm{d}}\sum_{d=1}^{N_\mathrm{d}}\sum_{h=1}^{24}\frac{|p_{d,h}-\hat{p}_{d,h}|}{|{p}_{d,h}|},
\end{align}

\noindent where $p_{d,h}$ and $\hat{p}_{d,h}$ respectively represent the real and forecasted price on day $d$ and hour $h$, and $N_\mathrm{d}$ is the number of days in the out-of-sample test period, i.e.\ in the test dataset.

Since absolute errors are hard to compare between different datasets, the MAE and RMSE are not always very informative. Moreover, since electricity costs and profits are often linearly dependent on the electricity prices, metrics based on quadratic errors, e.g.\ RMSE, are hard to interpret and do not accurately represent the underlying problem of most forecasting users. In particular, in most electricity trade applications, the underlying risk, profits, and costs depend linearly on the price and on the forecasting errors. Hence, linear metrics  represent better than quadratic metrics the underlying risks of forecasting errors.

Similarly, since MAPE values become very large with prices close to zero (regardless of the actual absolute errors), the MAPE is usually dominated by the periods of low prices and is also not very informative. While the \textit{symmetric mean absolute percentage error} (sMAPE) defined\footnote{Note, that there are multiple versions of sMAPE, here we consider the most sensible one according to \cite{Hyndman2014}.} as: 
\begin{equation}
\mathrm{sMAPE} = \frac{1}{24\,N_\mathrm{d}}\sum_{d=1}^{N_\mathrm{d}}\sum_{h=1}^{24}2\frac{|p_{d,h}-\hat{p}_{d,h}|}{|{p}_{d,h}| + |\hat{p}_{d,h}|}
\end{equation}
solves some of these issues, it has (as any metric based on percentage errors) a statistical distribution with undefined mean and infinite variance \cite{Hyndman2006}.

\subsubsection{Scaled errors}
In this context, several studies advocate for the use of scaled errors \cite{Weron2014,Hyndman2006,Hyndman2018}, where a scaled error is simply the MAE scaled by the in-sample MAE of a naive forecast. A scaled error has the nice interpretation of being lower/larger than one if it is better/worse than the average naive forecast evaluated in-sample. 

A metric based on this concept is the \textit{mean absolute scaled error} (MASE), and in the context of one-step ahead forecasting is defined as \cite{Hyndman2006}:

\begin{equation}
	\mathrm{MASE} = \frac{1}{N}\sum_{k=1}^{N}\frac{|p_k-\hat{p}_k|}{\frac{1}{n-1}\sum_{i=2}^{n} |p^\mathrm{in}_i - p^\mathrm{in}_{i-1} |},
\end{equation}

\noindent where $p^\mathrm{in}_i$ is the $i^\mathrm{th}$ price in the in-sample, i.e.\ training, dataset (note that in EPF $i=24d+h$), $p^\mathrm{in}_{i-1}$ is the one-step ahead naive forecast of $p^\mathrm{in}_i$, i.e.\ $\hat{p}^\mathrm{in}_i$, $N$ is the number of out-of-sample (test) datapoints, and $n$ the number of in-sample (training) datapoints. For seasonal time series, the MASE may be defined using the MAE of a seasonal naive model in the denominator \cite{Weron2014,Hyndman2018}. 

\subsubsection{Relative measures}
While scaled errors do indeed solve the issues of more traditional metrics, they have other associated problems that make them unsuitable in the context of EPF:

\begin{enumerate}
    \item As MASE depends on the in-sample dataset, forecasting methods with different calibration windows will naturally have to consider different in-sample datasets. As a result, the MASE of each model will be based on a different scaling factor and comparisons between models cannot be drawn.
    \item The same argument applies to models with and without rolling windows. The latter will use a different in-sample dataset at every time point while the former will keep the in-sample dataset constant.
    \item In ensembles of models with different calibration windows, the MASE cannot be defined as the calibration window of the ensemble is undefined.
    \item Drawing comparisons across different time series is problematic as electricity prices are not stationary. For example, an in-sample dataset with spikes and an out-of-sample dataset without spikes will lead to a smaller MASE than if we consider the same market but with the in-sample/out-sample datasets reversed.
\end{enumerate}

To solve these issues, we argue that a better metric is the \textit{relative} MAE (\rmae{}) \cite{Hyndman2006}. Similar to MASE, \rmae{} normalizes the MAE by the MAE of a naive forecast. However, instead of considering the in-sample dataset, the naive forecast is built based on the out-of-sample dataset. In the context of EPF, \rmae{} is defined as:

\begin{equation}
\mathrm{\rmae{}} = \frac{\displaystyle\frac{1}{24N_\mathrm{d}}\sum_{d=1}^{N_\mathrm{d}}\sum_{h=1}^{24}|p_{d,h}-\hat{p}_{d,h}|}{\displaystyle\frac{1}{ 24N_\mathrm{d}}\sum_{d=1}^{N_\mathrm{d}}\sum_{h=1}^{24} |p_{d,h} - \hat{p}^{\mathrm{naive}}_{d,h} |}, \label{eq:rmaeepf}
\end{equation}

\noindent where the $\frac{1}{24N_d}$ factor cancels out in the numerator and the denominator. There are three natural choices for the naive forecasts:
\begin{itemize}
\item $\hat{p}^{\mathrm{naive},1}_{d,h}=p_{d-1,h}$,
\item $\hat{p}^{\mathrm{naive},2}_{d,h}=p_{d-7,h}$,
\item $	\hat{p}^{\mathrm{naive},3}_{d,h} = 
	\begin{cases}
	p_{d-1,h}, &\, \text{if $d$ is Tue, Wed, Thu, or Fri,}\\
		p_{d-7,h}, &\, \text{if $d$ is Sat, Sun, or Mon.}		
	\end{cases}$
\end{itemize}

\noindent In the context of EPF, \rmae{} using $\hat{p}^{\mathrm{naive},2}_{d,h}=p_{d-7,h}$ is arguably the best choice for two reasons: (i) it is easier to compute than the one based on $\hat{p}^{\mathrm{naive},3}_{d,h}$ and, unlike the \rmae{} based on $\hat{p}^{\mathrm{naive},1}_{d,h}$, it captures weekly effects; (ii) given a set of forecasting models, the relative ranking of the accuracy of the models is independent from the naive benchmark used (see last paragraph of this subsection for an explanation). Hence, for the remainder of the article we will use \rmae{} to explicitly refer to the \rmae{} based on $\hat{p}^{\mathrm{naive},2}_{d,h}$.
It is important to note that, similar to \rmae{}, one could also define the \textit{relative} RMSE (rRMSE) by dividing the RMSE of each forecast by the RMSE of a naive forecast.

Since the dependence on the in-sample dataset is removed, using a rolling window is no longer a problem as the out-of-sample dataset stays the same. Similarly, models with different calibration windows can be compared and the \rmae{} of ensembles is properly defined. Moreover, as the metric is normalized by the MAE of a naive forecast for the same sample, the problem with drawing conclusions in non-stationary time series is mitigated. As before, we can also define the \rmae{} for seasonal time series:

Due to its better properties,  \rmae{} should always be used to evaluate new methods in EPF. In particular, while it can be used in conjunction with other metrics, it is important to include and employ \rmae{} to obtain more fair evaluations and comparisons. 

With that in mind, the accuracy of the open-access models in the open-access benchmark dataset is computed considering \rmae{}, sMAPE, MAPE, MAE, and RMSE. Then, an analysis of the different metrics is provided (see Section \ref{sec:metdisc}). Finally, the forecasts themselves are provided as csv files so that the accuracy results can be updated in case more adequate metrics are developed in the future.

As a final remark, let us to note that, given a set of forecasting models, the relative ranking of the accuracy of the models is independent from the naive benchmark used for the \rmae{}  or MASE. Changing it simply changes the denominator  but preserves the numerator, and since the change in the denominator is the same across all methods, the relative ranking is preserved. Furthermore, as the numerator is the MAE, it follows that the ranking based on the \rmae{} or MASE will be the same as that based on the MAE.

\subsection{Statistical testing}
While using adequate metrics to compare the accuracy of the forecasts is important, it is also necessary to analyze whether any difference in accuracy is statistically significant. This is paramount to conclude whether the difference in accuracy does really exist and is not simply due to random differences between the forecasts. Despite its importance, the use of statistical testing has been downplayed in the EPF literature \cite{Weron2014}. In particular, most publications only compare the accuracy in terms of an error metric and do not analyze the statistical significance of the accuracy differences. This trend needs to be corrected in order to compare forecasting approaches with statistical rigor. Particularly, new studies need to ensure that:

\begin{itemize}
\item Any new method is compared against well-established methods using a statistical test.
\item The forecasts of the proposed methods are provided as open-access datasets. This ensures that, when new models are proposed, the difference in accuracy with the published methods can be analyzed in terms of statistical testing.
\end{itemize}

To facilitate statistical testing, we include in the propose open-source \texttt{epftoolbox} library \cite{benchmarkwebsite,epftoolboxdoc} the two most widely used statistical tests in EPF, i.e.\ the Diebold-Mariano and the Giacomini-White tests.

\subsubsection{The Diebold-Mariano test}
The Diebold-Mariano (DM) test \cite{Diebold1995} is probably the most commonly used tool to evaluate the significance of differences in forecasting accuracy. It is an asymptotic \textit{z}-test of the hypothesis that the mean of the \textit{loss differential} series: 
\begin{equation}\label{eqn:DM}
\Delta^{\mathrm{A, B}}_{d,h} = L(\varepsilon^\mathrm{A}_{d,h}) - L(\varepsilon^\mathrm{B}_{d,h})
\end{equation}
is zero, where $\varepsilon^\mathrm{Z}_{d,h}=p_{d,h}-\hat{p}_{d,h}$ is the prediction error of model Z for day $d$ and hour $h$, and $L(\cdot)$ is the loss function. For point forecasts, we usually take $L(\varepsilon^\mathrm{Z}_{d,h})=|\varepsilon^\mathrm{Z}_{d,h}|^p$ with $p=1$ or $2$, which corresponds to the absolute and squared losses, respectively; for probabilistic forecasts, $L(\cdot)$ may be any strictly proper scoring rule, in particular the pinball loss, the \textit{continuous ranked probability score} (CRPS), or the energy score \cite{Nowotarski2018,wer:zie:20,gia:rav:ros:20}. Given the loss differential series, we compute the statistic:
\begin{equation}
\mathrm{DM} = \sqrt{N}\frac{\hat\mu}{\hat\sigma},
\end{equation}
where $\hat\mu$ and $\hat\sigma$ are the sample mean and standard deviation of $\Delta^{\mathrm{A, B}}_{d,h}$, respectively, and $N$ is the length of the out-of-sample test period. Under the assumption of covariance stationarity of $\Delta^{\mathrm{A, B}}_{d,h}$, the DM statistic is asymptotically standard normal, and one- or two-sided asymptotic tail probabilities can be easily computed.

It is important to note three things. Firstly, the DM test is model-free, i.e.\ it compares forecasts (of models), not models themselves. Secondly, although in the standard formulation \cite{Diebold1995} the DM test compares forecasts via the null hypothesis of the expected loss differential being zero, it is more informative to compute the $p$-values of two one-sided tests:
\begin{enumerate}
    \item with the null hypothesis $H_0:E(\Delta^{\mathrm{A, B}}_{d,h}) \le 0$, 
    \item with the alternative hypothesis null $H_1:E(\Delta^{\mathrm{A, B}}_{d,h}) \ge 0$. 
\end{enumerate}
The lower the $p$-value\footnote{Recall, that the $p$-value is the probability of obtaining results (in our case -- loss differentials) at least as large as the ones actually observed, assuming that the null hypothesis is correct.}, i.e.\ the closer it is to zero, the more the observed data is inconsistent with the null hypohtesis. If the $p$-value is less than the commonly accepted level of 5\%, the null hypothesis is typically rejected. In the DM test, this means that the forecasts of model B are significantly more accurate than those of model A. 

Thirdly, the DM test requires (only) that the loss differential be covariance stationary.\footnote{Actually
covariance stationarity is sufficient but may not be strictly
necessary \cite{die:15}.} This may not be satisfied by forecasts in day-ahead markets, since the predictions for all 24 hours of the next day are computed at the same time, using the same information set. Hence, following \cite{wer:zie:20}, we recommend two variants of the DM test in the context of day-ahead EPF:
\begin{itemize}
	\item a \textit{univariate} variant with 24 independent tests performed\footnote{We assume that a day-ahead market has 24 prices. For markets with prices every half hour, the univariate variant comprises 48 independent tests.}, one for each hour of the day, and comparisons based on the number of hours for which the predictions of one model are significantly better than those of another, i.e.\ the number of hours for which the null hypothesis is rejected, 
	\item a \textit{multivariate} variant with the test performed jointly for all hours using the `daily' or multivariate loss differential series:
    \begin{equation}\label{eq:CPA_eq}
    \Delta^{\mathrm{A, B}}_{d} = ||\varepsilon^\mathrm{A}_d||_p - ||\varepsilon^\mathrm{B}_d||_p,
    \end{equation}
    where $\varepsilon^\mathrm{Z}_d$ is the 24-dimensional vector of prediction errors of model Z for day $d$,  $||\varepsilon^\mathrm{Z}_d||_p = (\sum_{h=1}^{24} |\varepsilon^\mathrm{Z}_{d,h}|^p)^{1/p}$ is the $p$-th norm of that vector with $p=1$ or $2$.
\end{itemize}
\noindent The univariate version of the test has the advantage of providing a deeper analysis as it indicates which forecast is significantly better for which hour of the day  \cite{Bordignon2013,Nowotarski2014,Uniejewski2016,Lago2018a,Nowotarski2018,gia:rav:ros:20}. The multivariate version, introduced in \cite{Ziel2018},  enables a better representation of the results as it summarizes the comparison in a single $p$-value, which can be conveniently visualized using heat maps arranged as chessboards \cite{Uniejewski2018a,Uniejewski2018,Hubicka2019,Marcjasz2019}, see Figure \ref{fig:CPA}.

\subsubsection{The Giacomini-White test}

In some of the more recent EPF studies \cite{Marcjasz2018,Serafin2019,Marcjasz2020}, the DM test has been replaced by the Giacomini-White (GW) test \cite{Giacomini2006} for \textit{conditional predictive ability}. The latter is preferred because it can be regarded as a generalization of the DM test for \textit{unconditional predictive ability}: while both tests can be used for nested and non-nested models\footnote{This holds as long as the calibration window does not grow with the sample size \cite{Giacomini2013}. This is satisfied for rolling windows, but not for extended calibration windows.}, only the GW test accounts for parameter estimation uncertainty through `conditioning' \cite{wer:zie:20}.  

Like the DM test, also the GW test has two variants in day-ahead EPF -- the univariate and the multivariate. Without loss of generality, let us focus on the latter. It starts by building a multivariate loss differential series, see \eqref{eq:CPA_eq}, for a pair of forecasts (of models A and B).
Next, the test considers the following regression:   
\begin{equation}\label{eqn:GW}
\Delta^{\mathrm{A,B}}_d=\boldsymbol{\phi}' {X}_{d-1} + \epsilon_d, 
\end{equation}
where ${X}_{d-1}$ contains elements from the information set on day $d-1$, i.e.\ a constant and lags of $\Delta^{\mathrm{A, B}}_{d}$. Note that $\epsilon_d \ne \varepsilon^\mathrm{Z}_d$, i.e.\ $\epsilon_d$ is not the 24-dimensional vector of prediction errors for day $d$ and model $Z$ but simply an error term in the regression. Also note that using this notation the DM test can be written as \cite{Giacomini2013}:
\begin{equation}\label{eqn:DM:asGW}
\Delta^{\mathrm{A,B}}_d=\mu + \epsilon_d, 
\end{equation}
i.e.\ with ${X}_{d-1}$ containing just a constant. 
Finally, like for the DM test, to check the significance of differences in forecasting accuracy, the $p$-values of two one-sided tests can be computed.
The interpretation and possible visualization (see Figure \ref{fig:CPA}) are analogous to that of the DM test.



\subsection{Recalibration}
An issue with many EPF studies is that forecasting models are not recalibrated. Instead, they are often estimated once using the training dataset and directly evaluated in the whole test dataset. This is problematic as it does not represent real-life conditions where forecasting models are retrained (often on a daily basis) to account for the latest market information.

To have models that are evaluated in realistic conditions, they need to be retrained considering the new incoming flow of market information. As an example, for the day-ahead market, a forecasting model should be retrained on a daily basis as new information is available. Considering a testing period of a year, this means that a realistic evaluation requires estimating the forecasting model 365 times.

\subsection{Ex-ante hyperparameter optimization}
\label{sec:hyperopt}
A common issue in the current EPF literature is that the hyperparameter selection is often either done ex-post \cite{Anamika2018,Ibrahim2019,Panapakidis2016,Singh2015,Reddy2016,Peter2016} or its details are not sufficiently explained \cite{Ugurlu2018,Xu2019,Xie2018,Singh2018,Bisoi2018,Ghofrani2017,Lahmiri2017,Varshney2016,Xiao2017,Ebrahimian2018,Zhu2018}. As an example, when models based on neural networks are proposed, the details on how the number of neurons are selected are usually not provided. In other cases, while the approach is provided, it is often based on analyzing different configurations of neurons using the test dataset and selecting the one that works best, i.e.\ ex-post hyperparameter selection.

Not providing enough details on how hyperparameters are selected is an obvious problem as it prevents reproducing research. Similarly, performing hyperparameter optimization ex-post leads to overfitting the test dataset, i.e.\ the model is partially optimized using the same dataset used for evaluating the model, and it grants the model an unfair and non-existent advantage over other models.

To prevent this, the selection of hyperparameters should be explicitly explained and always performed ex-ante using a validation dataset. With that motivation, for the open-access methods proposed, not only do we explain how the hyperparameters are obtained, but we also provide  within the toolbox \cite{benchmarkwebsite,epftoolboxdoc} a module for hyperparameter selection and the files containing the results of the hyperparameter optimization of the current study.

\subsection{Computation time}
\label{sec:compttime}
An even more common problem is the fact that new models are very rarely compared in terms of their computational requirements \cite{Anamika2018,Bento2018,Khajeh2017,Varshney2016,Xiao2017,Ebrahimian2018,Kim2015,Hong2014a,Abedinia2015,Pourdaryaei2019,Nazar2018,Itaba2017,Olamaee2016,Zhou2018,Zhang2019,Yang2017,Zhu2018}. Although a model might be marginally better than another, it might not be worthwhile to deploy it in a practical application if its computational requirements are much larger. Particularly, higher computational requirements might pose two problems: 

\begin{enumerate}
\item As mentioned before, forecasting models should ideally be recalibrated on a daily basis. Hence, a forecasting method is only suitable if its computational time allows this recalibration to take place. In this context, the maximum available time for estimating a model will depend on each electricity market but, as a rule of thumb, it can be argued that any model that requires more than 30\,min or 1\,h will unlikely be suitable for forecasting prices in the spot markets.
\item Besides recalibration, the second issue with computation time is its cost. If the computational capabilities are too large, the benefits of using a marginally better forecast might be lower than the cost of running the forecasting model on a much more expensive computer.
\end{enumerate}

Hence, when new forecasting models are proposed, we argue that it is very important to provide their computation times. Moreover, we also argue that for a model to be better than the existing methods, it does not necessarily have to be the most accurate one. Instead:

\begin{enumerate}
    \item If its computational time is large, i.e.\ in the order of minutes, the model should indeed be more accurate than all state-of-the-art models, e.g.\ DNNs.
    \item If its computational time is small, i.e.\ in the order of seconds, the model should be more accurate than the state-of-the-art models with low computational requirements, e.g.\ LEAR.
\end{enumerate}

In this article, we provide an analysis of the computational requirements of the proposed open-access models so other researchers can easily make such comparisons.

\subsection{Reproducibility}
Another related issue is that some studies lack enough details to replicate the research. Missing details vary from study to study but the four most common are:

\begin{enumerate}
	\item the dataset used for testing and evaluation is not defined \cite{Aggarwal2017,Hong2014a,Talari2017,Singh2017,Khan2017,Afrasiabi2019a,Zhu2018};
		\item the dataset used for training is not defined \cite{Bento2018,Khajeh2017,Singh2018,Talari2017,Khan2017};
	\item the inputs of the model are unclear \cite{Wang2017a,Khan2017,Shrivastava2014,Jiang2015,Afrasiabi2019a};
	\item the selection of hyperparameters is unclear \cite{Ugurlu2018,Xu2019,Xie2018,Singh2018,Bisoi2018,Ghofrani2017,Lahmiri2017,Varshney2016,Xiao2017,Ebrahimian2018,Zhu2018}.
\end{enumerate}

\noindent To correct this, future EPF papers should provide enough details to allow replication and reviewers should verify that all necessary details of the employed datasets are always provided.

\subsection{Data contamination}
Another recurrent issue in the EPF literature  is data contamination, which appears when part of the training dataset is used for testing. Particularly, when working with time series data the test dataset should always comprise the last part of the dataset to avoid data contamination. If this is not done, the models can overfit the testing dataset and their accuracy can be overestimated. 

Despite the importance of correctly separating the training/validation dataset from the testing dataset, some studies in EPF:

\begin{enumerate}
	\item Do not specify the split between the training, validation, and test datasets \cite{Aggarwal2017,Hong2014a,Talari2017,Singh2017,Khan2017,Bento2018,Khajeh2017,Singh2018,Afrasiabi2019a,Zhu2018}. If the datasets are not specified, it is not possible to know whether data contamination occurs.
	\item Randomly sample the test dataset from the full dataset \cite{Nascimento2019,Kotur2016,Monteiro2016,Monteiro2015}, e.g.\ in a dataset comprising a year of data randomly selecting 4 weeks for testing and the remaining data for training.
	\item Have a partial or total overlap between the training/validation dataset and the testing dataset \cite{Anamika2016,Panapakidis2016,Anamika2018,Ibrahim2019}, e.g.\ by performing hyperparameter optimization ex-post.
\end{enumerate}

To correct this issue, it is important that any future research in EPF ensures that: 1) the split between the datasets is correctly described; 2) the test dataset does never overlap with the training or validation datasets; 3) the test dataset is always selected as the last segment of the full dataset.

\subsection{Software toolboxes}
A less pressing yet relevant issue is the use of state-of-the-art software toolboxes. When comparing new methods with methods from the literature, the latter should be modeled using adequate toolboxes. Particularly, it is important to use toolboxes that are continuously updated as implementing methods using outdated libraries leads to unfair evaluations. 

For example, in the context of neural networks, there are several open-source state-of-the-art toolboxes \cite{Chollet2015} that are continuously updated and that grant access to the latest development in the field of DL. Yet, in the context of EPF, new methods are often compared with neural networks that are modeled using the \texttt{MATLAB} toolbox \cite{Bento2018,Khajeh2017,Peter2016,Varshney2016,Wang2017a,Xiao2017,Kim2015,Hong2014a,Kotur2016}, a toolbox that for many years was outdated and did not include many of the neural network developments that are critical in EPF, e.g.\ state-of-the-art activation functions or stochastic gradient descent algorithms \cite{Lago2018a}. As a result, many of the existing comparisons in EPF are based on evaluations where the accuracy of neural networks might be underestimated. 

Besides using state-of-the-art software toolboxes, e.g.\ the \texttt{python} library \texttt{keras} for deep learning, it is also important to employ (whenever possible) free-to-access libraries so that research can be replicated by anyone.

\subsection{Combining forecasts}
\label{sec:ensemble}
As a final guideline, it is important to indicate the importance of ensembles in the context of EPF. In general, although exceptions exist \cite{Atiya2020}, combining different models leads to a higher accuracy \cite{Nowotarski2014,Marcjasz2018} and it is thus a good idea to build forecasts based on multiple models. However, as even the arithmetic average improves the accuracy of individual models, new ensemble techniques should be studied in comparison with other ensemble techniques, i.e.\ as done in \cite{Nowotarski2014}, and not simply w.r.t.\ the individual models.

{To maximize the forecasting accuracy, it is important to employ diverse forecasts \cite{Atiya2020}, e.g.\ forecasts generated using different data or different models. For EPF, the former can be achieved by considering models trained using different calibration window lengths  \cite{Hubicka2019,Serafin2019} and the latter using different modeling techniques or different sets of hyperparameters. To further maximize the performance, the number of models used in the ensemble should be limited \cite{Atiya2020}, e.g.\ 4--10, especially in the case of heavy-tailed data for which large ensembles tend to contain outliers more often, resulting in less accurate forecasts.

With that in mind, as part of the open-access benchmark and toolbox \cite{benchmarkwebsite,epftoolboxdoc}, we also propose a series of simple ensemble techniques. Particularly, as explained in Section \ref{sec:Models}, we provide an ensemble of four LEAR models that are estimated over different calibration windows and combined using a simple arithmetic average and another ensemble using four DNNs that are estimated for different hyperparameters and combined using the arithmetic average.

\section{Evaluation of state-of-the-art methods}
\label{sec:eval}
In this section, we present the results of the open-source benchmark methods for all five datasets. For the sake of clarity, we divide the section into two parts respectively comprising the results for the error metrics and the results for statistical testing. 

\subsection{Accuracy metrics}
\label{sec:benchmark_accuracy}
We first start by presenting the results of the open-access benchmark models 
in terms of accuracy metrics. 

\subsubsection{Individual models}
Table \ref{tab:compindiv} compares the performance of the two individual models and their 4 variations in terms of \rmae{}, MAE, MAPE, SMAPE, and RMSE. The LEAR model is displayed for 4 different calibration windows representing 56, 84, 1092, and 1456 days, i.e.\ 8 weeks, 12 weeks, 3 years, and 4 years. The four DNNs are obtained by performing the hyperparameter/feature optimization process four times and using the best hyperparameter/feature selection of every run (see Sections \ref{sec:hyperoptDNN} and \ref{sec:ensemblesintro} for further details)\footnote{Note that, for the sake of simplicity, the features and hyperparameter selection for each model are not provided. However, they can be obtained from the website \cite{benchmarkwebsite} accompanying this study}. Several observations can be made:

\begin{itemize}
	\item The MAPE seems an unreliable metric as it completely disagrees with the other three linear metrics and the quadratic metric. In particular, while the \rmae{}, MAE, and sMAPE agree on what the best model is in all the cases, the MAPE almost never does so. This unreliability can be further seen in the German market: while the MAPE and sMAPE metrics usually have similar orders of magnitude, in the case of the German market the MAPE is approximately 10 times larger. This effect is due to prices in Germany being negative and very close to 0, leading in turn to very large MAPE values that bias the avarege MAPE.	
	\item The DNN models seem to be more accurate than the LEAR models. Particularly, in terms of linear metrics, the four DNN models perform better than all four LEAR models for the 5 datasets. 
	\item Although the RMSE displays slighly different results, this is expected as the metric is based on quadratic errors and not linear ones. Nonetheless, while the RMSE does display slightly different results, it still shows the superiority of the DNN model: even though the DNN is estimated to minimized absolute errors (unlike LEAR), the DNN is better in 4 of the 5 datasets. Moreover, even though the DNN seems to be worse in two markets, the RMSE metric does not correctly represent the underlying problem (see Sections \ref{sec:metricsguidelines} and \ref{sec:linearvssquare}) and it can be argued that it is not the best metric to assess the performance of EPF models. 
\end{itemize}

\begin{table*}[!ht]
	\centering
	\def\arraystretch{1.3}%
	\setlength{\tabcolsep}{6pt}
	\caption{Comparison between the two individual state-of-the-art open-source methods in terms of \rmae{}, MAE, MAPE, sMAPE, and RMSE. Each of the two methods is listed for four different configurations. The gray cells represent the best model for a given metric.}
		\newcommand{\cgray}{\cellcolor[gray]{0.85}}
	\newcommand{\bestmodel}{\cellcolor[gray]{0.85}}
	\begin{tabular}{ll|cccccccc}
		& & \bfseries{DNN$_{1}$} & \bfseries{DNN$_{2}$} & \bfseries{DNN$_{3}$} & \bfseries{DNN$_{4}$} & \bfseries{LEAR$_{56}$} & \bfseries{LEAR$_{84}$} & \bfseries{LEAR$_{1092}$} & \bfseries{LEAR$_{1456}$} \\
		\hline
		\multirow{4}{*}{\bfseries{NP}} %
         & \bfseries{\rmae{}} & 0.471 & \cgray 0.415 & 0.437 & 0.438 & 0.475 & 0.472 & 0.482 & 0.481 \\
         & \bfseries{MAE} & 1.946 & \cgray 1.717 & 1.808 & 1.812 & 1.964 & 1.952 & 1.993 & 1.990 \\
         & \bfseries{MAPE\,[\%]} & 6.04 & \cgray5.46 & 5.93 & 5.85 & 6.34 & 6.36 & 6.10 & 6.14 \\
         & \bfseries{sMAPE\,[\%]}&5.59 & \cgray 5.00 & 5.22 & 5.26 & 5.66 & 5.62 & 5.64 & 5.66\\		
         & \bfseries{RMSE} & 3.579  & \cgray 3.341  & 3.502  & 3.596  & 3.671    & 3.664    & 3.605      & 3.604 \\
         \hline
		\multirow{4}{*}{\bfseries{PJM}} %
         & \bfseries{\rmae{}} & 0.475 & 0.475 & 0.473 & \cgray 0.467 & 0.550 & 0.548 & 0.490 & 0.489 \\
         & \bfseries{MAE} & 3.005 & 3.008 & 2.995 & \cgray 2.956 & 3.477 & 3.467 & 3.098 & 3.095 \\
         & \bfseries{MAPE\,[\%]} & \cgray 28.87 & 29.74 & 29.87 & 29.10 & 32.52 & 32.34 & 30.28 & 30.24 \\
         & \bfseries{sMAPE\,[\%]} & 11.99 & 11.93 & 11.89 & \cgray 11.81 & 13.68 & 13.58 & 12.33 & 12.54 \\
        & \bfseries{RMSE} & 5.121  & 5.333  & 5.023  & \cgray 4.820   & 5.718    & 5.709    & 5.264      & 5.142    \\
		\hline
        \multirow{3}{*}{
             \textbf{EPEX}} %
& \bfseries{\rmae{}} & 0.608 & 0.600 & \cgray 0.597 & 0.608 & 0.682 & 0.669 & 0.649 & 0.653 \\
         \multirow{3}{*}{
             \textbf{BE}}        & \bfseries{MAE} & 6.181 & 6.094 & \cgray 6.066 & 6.173 & 6.924 & 6.798 & 6.594 & 6.634 \\
         & \bfseries{MAPE\,[\%]} & 24.83 & 28.69 & 24.08 & 30.46 & 32.88 & 32.34 & 26.26 & \cgray 22.64 \\
         & \bfseries{sMAPE\,[\%]} & 14.40 & 14.35 &\cgray 13.87 & 14.25 & 16.20 & 15.95 & 16.87 & 17.29 \\
         & \bfseries{RMSE} & 16.577 & \cgray 15.879 &  16.304 & 16.488 & 16.371   & 16.291   & 16.458     & 16.420 \\
         \hline
		\multirow{3}{*}{\textbf{EPEX}} %
         & \bfseries{\rmae{}} & 0.576 & 0.572 & \cgray 0.562 & 0.585 & 0.638 & 0.624 & 0.580 & 0.597 \\
         \multirow{3}{*}{
             \textbf{FR}}& \bfseries{MAE} & 4.223 & 4.193 & \cgray 4.118 & 4.292 & 4.681 & 4.575 & 4.250 & 4.378 \\
         & \bfseries{MAPE\,[\%]} & 15.75 & 16.52 & 15.13 & 15.55 & 19.03 & 18.09 & 14.95 & \cgray 14.90 \\
         & \bfseries{sMAPE\,[\%]} & 12.06 & 12.03 & \cgray 11.65 & 11.96 & 13.43 & 13.28 & 13.25 & 14.05 \\
        & \bfseries{RMSE} & 12.036 & 11.850  & 11.414 & 12.455 & 11.732   & \cgray 10.759   & 11.337     & 11.462    \\
		\hline
		\multirow{3}{*}{\textbf{EPEX}} %
         & \bfseries{\rmae{}} & \cgray0.395  & 0.398 &\cgray 0.395  & 0.413         & 0.469    & 0.458    & 0.431      & 0.437  \\
        \multirow{3}{*}{
             \textbf{DE}} & \bfseries{MAE} & \cgray3.601  & 3.633 & 3.605  & 3.771       & 4.283    & 4.18     & 3.93       & 3.988  \\
         & \bfseries{MAPE\,[\%]} & 103.14 & \cgray83.1  & 100.59 & 106.38       & 133.38   & 115.61   & 123.39     & 120.24    \\
         & \bfseries{sMAPE\,[\%]} & 14.84  & 15.06 &\cgray 14.74  & 15.28       & 16.54    & 16.27    & 16.79      & 17.15  \\
         & \bfseries{RMSE} & 6.256  &\cgray 6.135 & 6.25   & 6.324        & 7.713    & 7.397    & 6.526      & 6.502   
	\end{tabular}
 
	\label{tab:compindiv}
\end{table*}

\subsubsection{Ensembles}
The results for the ensemble methods are listed in Table \ref{tab:compens}, which compares the performance of the two ensemble models and the best DNN and LEAR models in terms of the rMAE metric, i.e.\ arguably the most reliable metric. From the table, several observations can be made:

\begin{itemize}
	\item As already argued in Section \ref{sec:ensemble}, combining models usually improves the accuracy. Particularly, the ensemble of DNNs is better than the best individual DNN model for all five markets and for all reliable metrics. Similarly, the ensemble of LEAR models is better than the best individual LEAR model for all markets and reliable metrics. The exception to this observation are the MAPE and RMSE metrics but, as already noted, MAPE is an unreliable metric and RMSE does not correctly represent the underlying problem of EPF. 
	\item As before, in terms of rMAE, the ensemble of DNNs is the most accurate model across all markets, which again seems to suggest that the DNN models are more accurate than the LEAR models.
\end{itemize}

\begin{table*}[ht]
	\centering
	\def\arraystretch{1.3}%
	\caption{Comparison between the ensembles of the state-of-the-art open-source methods in terms of \rmae{}, MAE, MAPE, and sMAPE. The comparison also includes, for each market, the best individual performing DNN and LEAR model in terms of \rmae{} and MAE, i.e.\ the two most reliable metrics. The gray cells represent the best model for a given metric.}
	\label{tab:compens}		
			\newcommand{\bestmodel}{\cellcolor[gray]{0.85}}
	\begin{tabular}{ll|cccc}
		&  & \textbf{DNN Ensemble} & \textbf{LEAR Ensemble} & \textbf{Best\footnote{Best in terms of \rmae{}/MAE.} DNN} & \textbf{Best LEAR} \\
		\hline
		\multirow{4}{*}{\textbf{NP}}  %
& \bfseries{\rmae{}}   & \bestmodel 0.403        & 0.420           & 0.415  & 0.472      \\
& \textbf{MAE}    & \bestmodel 1.667        & 1.738          & 1.717  & 1.952      \\
& \textbf{MAPE [\%]}   & \bestmodel 5.38         & 5.53           & 5.46   & 6.36       \\
& \textbf{sMAPE [\%]}  & \bestmodel 4.85         & 5.01           & 5.00      & 5.62       \\
& \textbf{RMSE} & \bestmodel 3.333        & 3.362 & 3.341 & 3.604 \\
		\hline
		\multirow{4}{*}{\textbf{PJM}} %
 & \bfseries{\rmae{}}   & \bestmodel0.439        & 0.476          & 0.467  & 0.489      \\
 & \textbf{MAE}    &\bestmodel 2.779        & 3.013          & 2.956  & 3.095      \\
 & \textbf{MAPE [\%]}   & \bestmodel28.66        & 30.13          & 29.10   & 30.24      \\
 & \textbf{sMAPE [\%]}  & \bestmodel11.22        & 11.98          & 11.81  & 12.54      \\
 & \textbf{RMSE} & \bestmodel 4.637  & 5.127 & 4.820 & 5.142 \\
 \hline
\multirow{3}{*}{\textbf{EPEX}}  %
& \bfseries{\rmae{}}   & \bestmodel0.573        & 0.604          & 0.597  & 0.649      \\
 \multirow{3}{*}{\textbf{BE}}& \textbf{MAE}    & \bestmodel5.821        & 6.140           & 6.066  & 6.594      \\
 & \textbf{MAPE [\%]}   & 26.11        & \bestmodel20.72          & 24.08  & 26.26      \\
 & \textbf{sMAPE [\%]}  & \bestmodel13.33        & 14.55          & 13.87  & 16.87      \\
 & \textbf{RMSE}  & 16.127       & 15.974 & \bestmodel  15.879 & 16.371 \\
		\hline
		\multirow{3}{*}{\textbf{EPEX}} %
& \bfseries{\rmae{}}   & \bestmodel0.533        & 0.543          & 0.562  & 0.58       \\
 \multirow{3}{*}{\textbf{FR}}& \textbf{MAE}    & \bestmodel3.910         & 3.980           & 4.118  & 4.250       \\
 & \textbf{MAPE [\%]}   & 14.77        &\bestmodel 14.68          & 15.13  & 14.95      \\
 & \textbf{sMAPE [\%]}  & \bestmodel10.98        & 11.57          & 11.65  & 13.25      \\
 & \textbf{RMSE} & 11.738       & \bestmodel 10.676 & 11.414 & 10.759 \\
		\hline
		\multirow{3}{*}{\textbf{EPEX}} %
& \bfseries{\rmae{}}   & \bestmodel 0.377        & 0.395          & 0.395    & 0.431       \\
\multirow{3}{*}{\textbf{DE}} & \textbf{MAE}    & \bestmodel 3.441        & 3.609          & 3.601    & 3.93     \\
 & \textbf{MAPE [\%]}   & 95.76        & 113.98         & \bestmodel 83.1     & 115.61    \\
 & \textbf{sMAPE [\%]}  &  \bestmodel  14.19        & 14.74          & 14.74    & 16.27    \\
 & \textbf{RMSE} &\bestmodel  5.997        & 6.508          & 6.135    & 6.502      \\
 \end{tabular}

\end{table*}

\subsection{Statistical Testing}
\label{sec:benchmark_test}
In this section, we present the results of the open-access benchmark models in terms of the statistical tests. For the sake of simplicity, we present together the results for individual methods and ensembles. The results are based on the multivariate GW test using the $L_1$ norm in \eqref{eq:CPA_eq}, i.e.\ with the following loss differential series: 
\begin{equation}\label{eqn:GW:L1}
\Delta^{\mathrm{A, B}}_{d} = \textstyle\sum_{h=1}^{24} |\varepsilon^\mathrm{A}_{d,h}| - \sum_{h=1}^{24} |\varepsilon^\mathrm{B}_{d,h}|.
\end{equation}
While squared losses could also be used, we do not consider them here because absolute errors better represent the underlying problem in EPF, see Section \ref{sec:linearvssquare} for a discussion.

In Figure \ref{fig:CPA} we display the results for the five markets. More precisely, we use heat maps arranged as chessboards to indicate the range of the obtained $p$-values. The closer they are to zero (dark green) the more significant
is the difference between the forecasts of a model on the X-axis
(better) and the forecasts of a model on the Y-axis (worse). For instance, for the EPEX-DE market the first row is green indicating that the forecasts of LEAR$_{56}$ are significantly
outperformed by those of all other models. We can observe that:

\begin{figure*}[ht]
\begin{subfigure}[t]{0.325\textwidth}
\begin{center}
	\includegraphics[width=\textwidth]{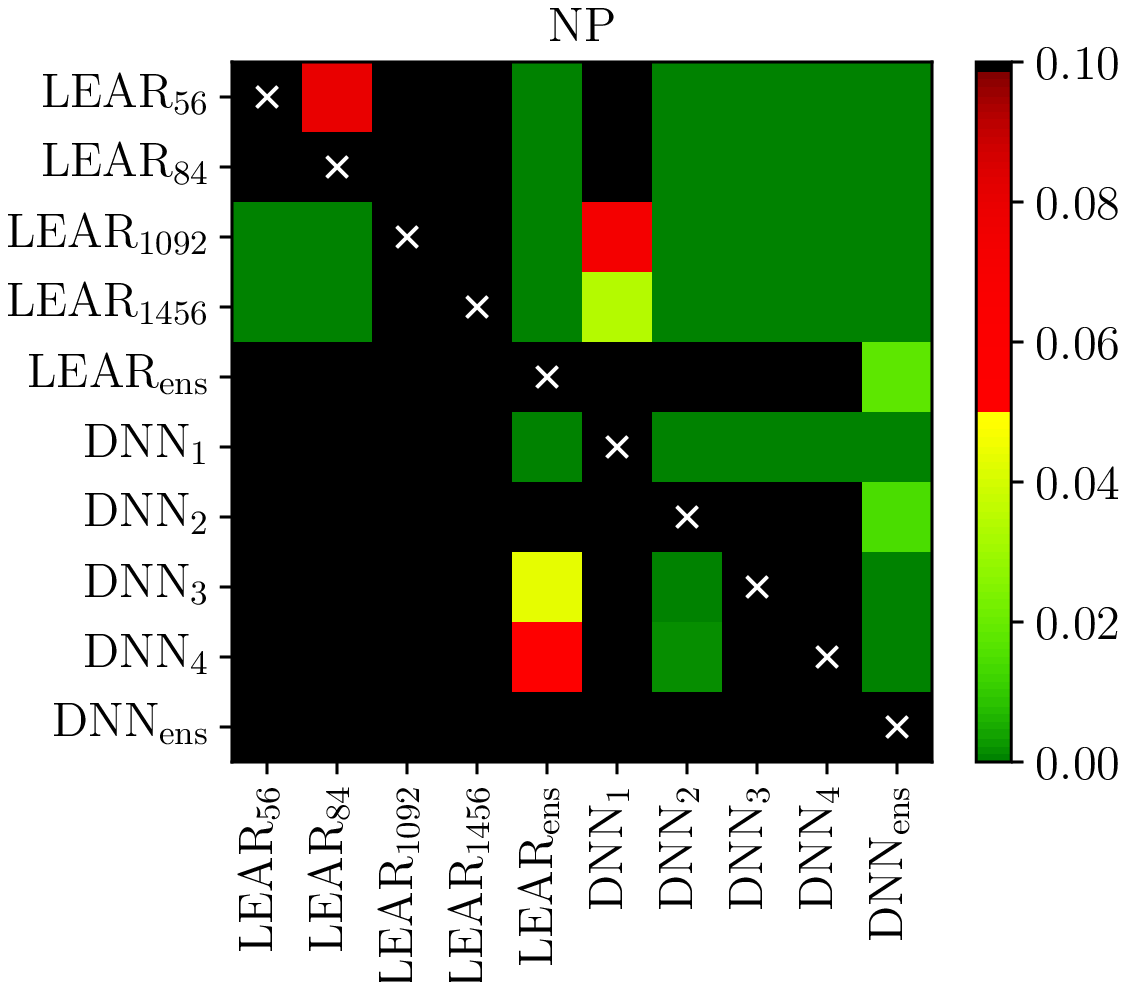}
\end{center}
\end{subfigure}
\begin{subfigure}[t]{0.325\textwidth}
\begin{center}
	\includegraphics[width=\textwidth]{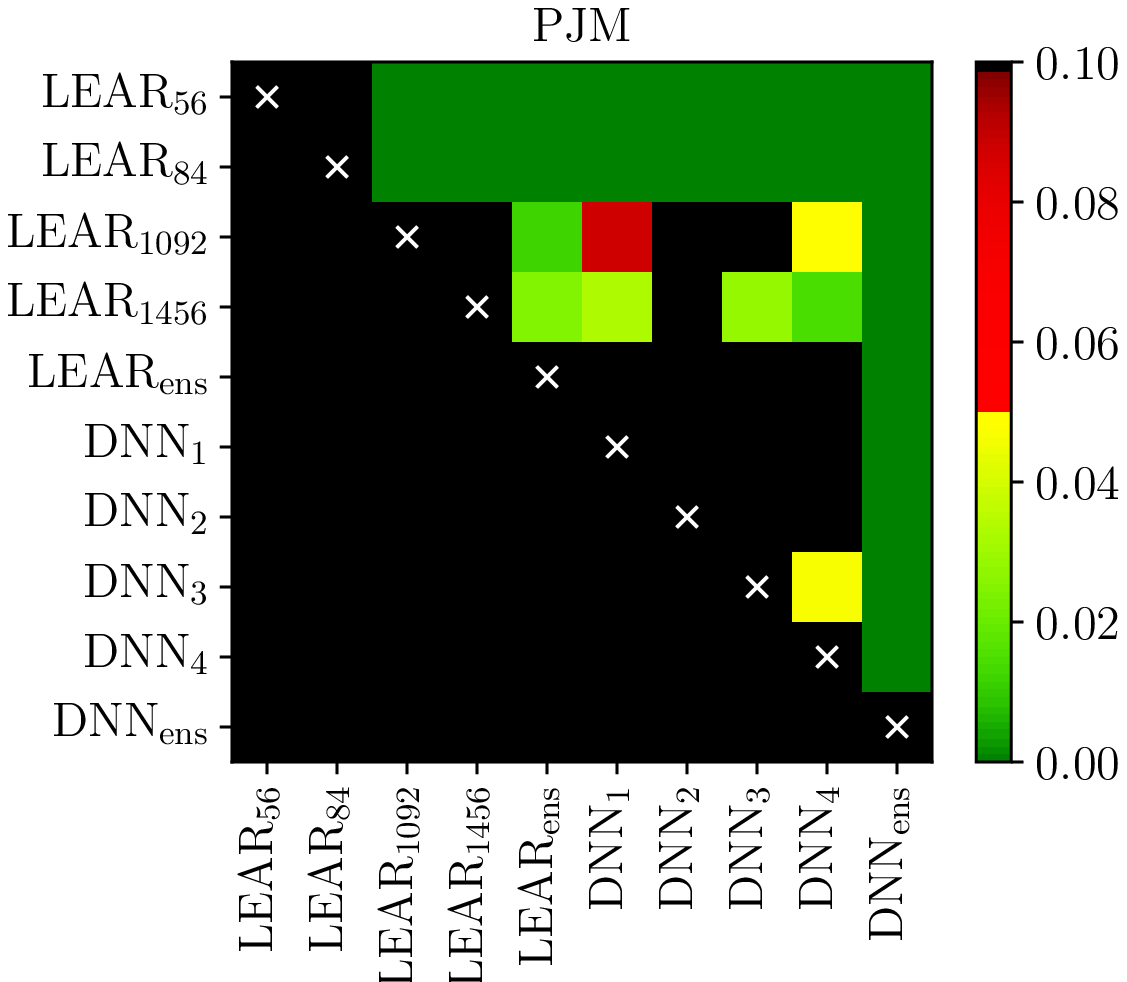}
\end{center}
\end{subfigure}
	\begin{subfigure}[t]{0.325\textwidth}
		\begin{center}
			\includegraphics[width=\textwidth]{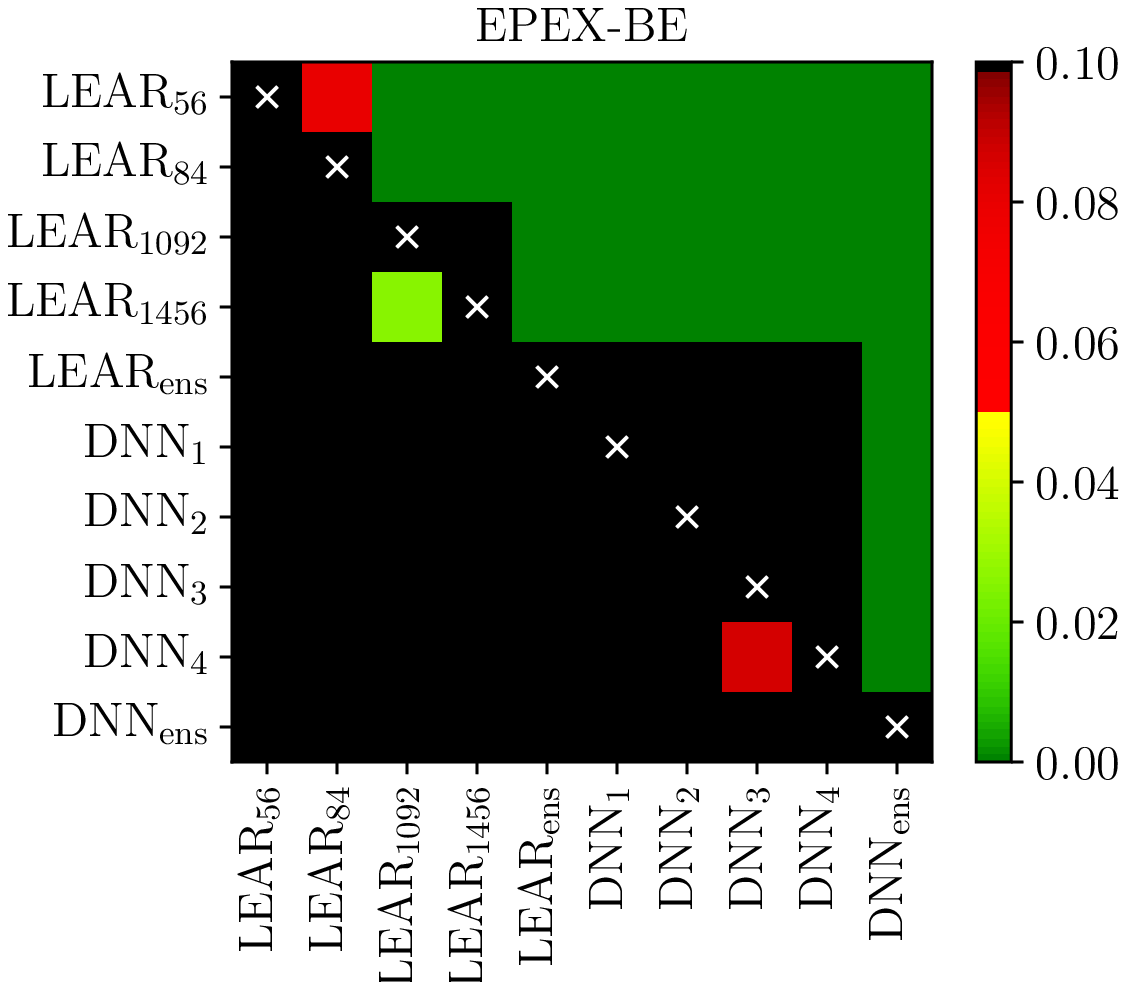}
		\end{center}
	\end{subfigure}
	\begin{subfigure}[t]{0.16\textwidth}~~~\end{subfigure}
	\begin{subfigure}[t]{0.32\textwidth}
		\begin{center}
			\includegraphics[width=\textwidth]{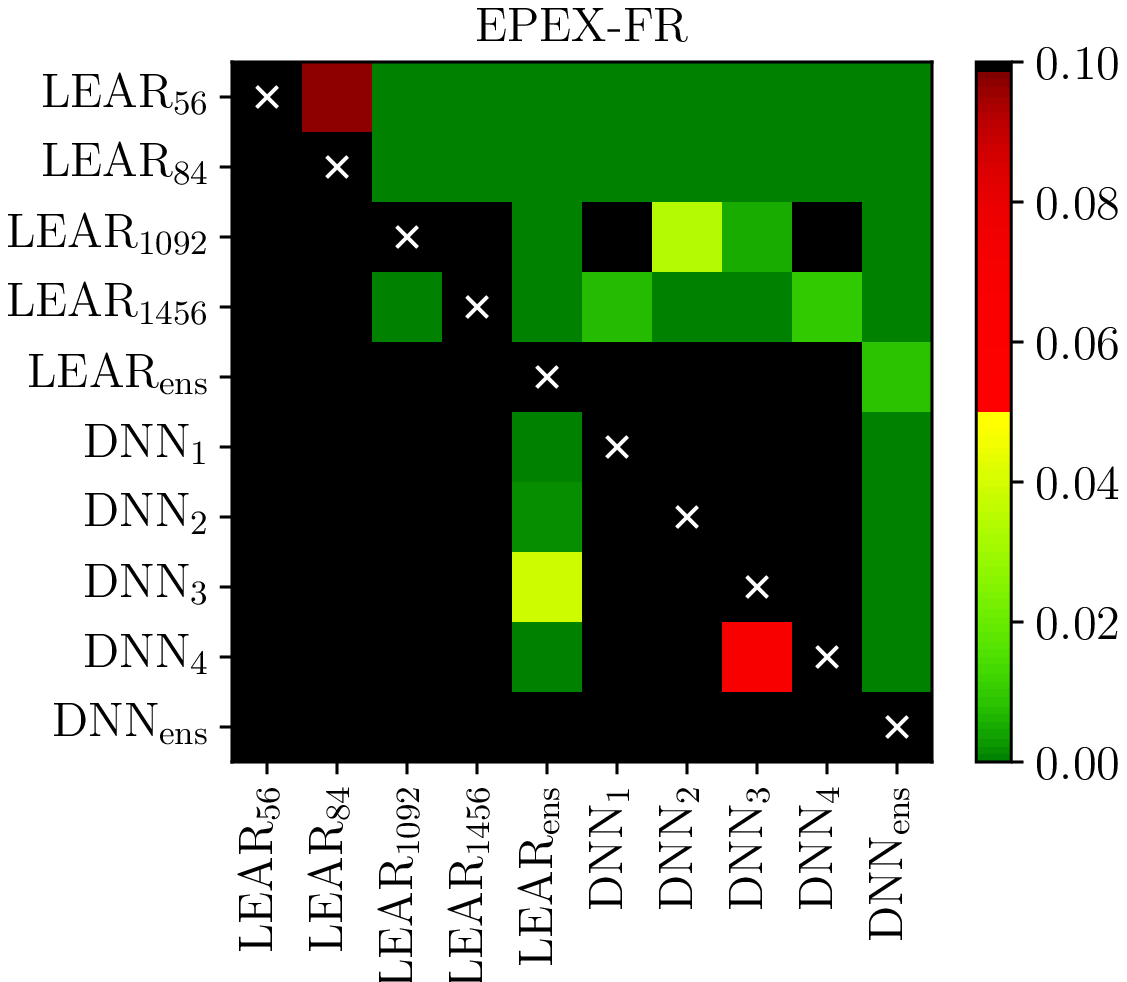}
		\end{center}
	\end{subfigure}
	\begin{subfigure}[t]{0.32\textwidth}
		\begin{center}
			\includegraphics[width=\textwidth]{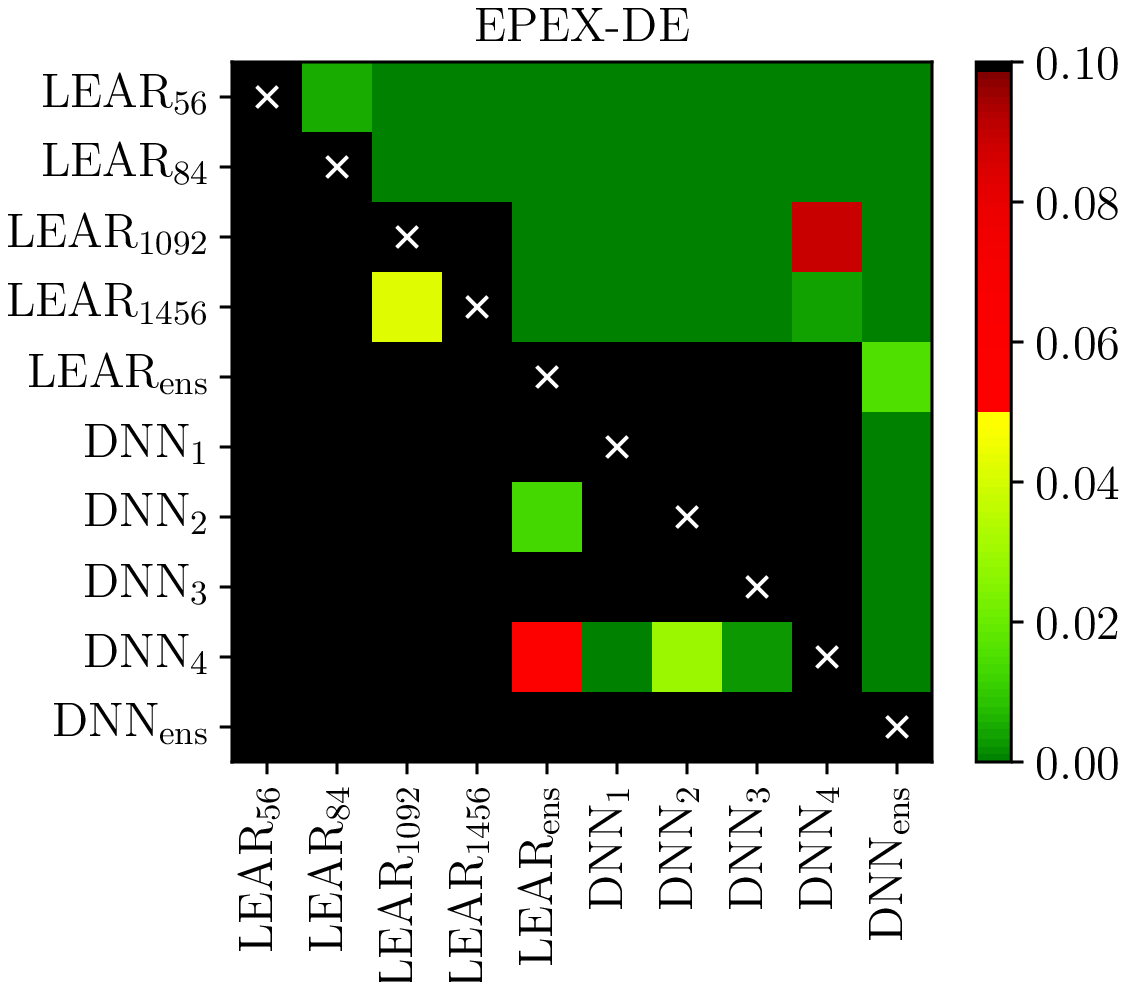}
		\end{center}
	\end{subfigure}
\caption{Results of the GW test with the multivariate loss differential series \eqref{eqn:GW:L1} for the eight individual models and the two ensembles. A heat map is used to indicate the range of the obtained $p$-values for each of the five markets. The closer the p-values are to zero (dark green), the more significant the difference is between the forecasts of a model on the X-axis (better) and the forecasts of a model on the Y-axis (worse). Black color indicates p-values above the color map limit, i.e.\ p-values larger or equal than 0.10.}
\label{fig:CPA}
\end{figure*}

\begin{itemize}
\item For all markets the last column is green indicating
that the forecasts of the ensemble of DNNs are statistically significantly better than the predictions of all the other models for all 5 datasets. 
	\item The forecasts of LEAR$_\mathrm{ens}$ are statistically significantly better than those of all individual LEAR models.	Together with the previous observation, i.e.\ the superiority of the DNN ensemble, this shows that the predictions of ensemble models usually improve upon the forecasting accuracy of individual methods.
	\item In one dataset (EPEX-BE), the forecasts of all the individual DNN methods are statistically significantly better than those of the individual LEAR models. In three other datasets (NP, EPEX-FR, and EPEX-DE), the forecasts of all the individual DNN methods are statistically significantly better than 3 out of the 4 individual LEAR models. For all datasets, there is at least 1 DNN model whose forecasts are statistically significantly better than those of all LEAR models.
	\item The forecasts of the individual LEAR models are never significantly better than those of the individual DNN models. Overall, it seems that forecasts based on DNNs are more likely to obtain significantly better results; this is particularly true for the DNN ensemble.
\end{itemize}

\subsection{Computation time}
\label{sec:benchmark_time}
As described in Section \ref{sec:compttime}, besides comparing the predictive accuracy, it is also necessary to analyze the computation time of the forecasting methods. Table \ref{tab:computation_time} lists a comparison of the computation time required for estimating the models considered, i.e.\ the time required to recalibrate each model on a daily basis. As the computation time is non-deterministic, its value is given as a range. These data were obtained using a regular laptop quadcore CPU, i.e.\ the i7-6920HQ.

\begin{table}[ht]
    \caption{Computation time that each benchmark model requires to perform a daily recalibration.}
    \label{tab:computation_time}
    \centering
    \begin{tabular}{c|c}
         & \textbf{Time}\\
         \hline
      \bfseries LEAR & 1--10 seconds  \\
     \bfseries LEAR Ensemble & 20--25 seconds  \\    
   \bfseries   DNN & 2--5 minutes  \\
   \bfseries DNN Ensemble & 8--20 minutes  \\    
    \end{tabular}
\end{table}

As can be observed, although the LEAR model performs slightly worse than the DNN model, its computation time is 30 to 100 times lower; particularly, when considering the maximum computation time of both methods, the LEAR model is 50 times faster.


\subsection{Discussion and remarks}
\label{sec:disc}
In this section, we provide some final remarks behind the motivation of the metrics employed, we briefly analyze the influence of the different metrics considered, and provide a discussion on comparing new models.

\subsubsection{Absolute vs.\ squared errors}
\label{sec:linearvssquare}

Throughout the text, we have mostly considered accuracy metrics based on absolute/linear errors, i.e.\ metrics that evaluate the accuracy of predicting the median of the distribution. Since the LEAR model is estimated by minimizing squared errors, thus leading to forecasts of the mean \cite{Hyndman2018}, one could argue that a metric/test based on squared errors should be preferred. While the argument has some merits, we focused on absolute metrics for three reasons:

\begin{itemize}
    \item The metric used to evaluate the accuracy should be the one that better represents the underlying problem. In the case of EPF,  since the cost of purchasing electricity is linear, linear metrics are arguably the best to quantify the risk associated with forecasting errors.
    \item While we provided the RMSE results, they are qualitatively the same as for MAE/rMAE. Hence, as absolute errors better represent the underlying problem of EPF and the results are similar, the RMSE results are not analyzed here in detail due to space limitations. 
    \item While the LEAR model is indeed estimated using squared errors, this is partly done because the techniques to efficiently estimate the LASSO, e.g.\ coordinate descent, are based on square errors. This gives the LEAR model a computational advantage over the DNN. An alternative would be to use regularized quantile regression \cite{li:zhu:08} leading, however, to an increased computational burden with little benefits on the accuracy in terms of MAE/rMAE.
\end{itemize}


\subsubsection{Metrics}
\label{sec:metdisc}
The obtained results validate the general guideline proposed in Section \ref{sec:metricsguidelines} regarding accuracy metrics: research in EPF should avoid MAPE and only use metrics like sMAPE or RMSE in conjunction with any version of \rmae{}. Particularly, the results validate the following three claims:

\begin{itemize}
	\item MAE is as reliable as \rmae{}. However, as the errors are not relative, comparison between datasets is not possible and \rmae{} is preferred.	
	\item sMAPE is more reliable than MAPE and it agrees with MAE/\rmae{}. Yet, it has the problem of an undefined mean and an infinite variance. Thus, it is less reliable than \rmae{}.	
	\item MAPE is not a reliable metric as it gives more importance to datapoints close to zero. As such, using MAPE can lead to misleading results and wrong conclusions.
	\item RMSE is more reliable than MAPE but it does not represent correctly the underlying risks of EPF. Hence, it should not be used alone to evaluate forecasting models. 
\end{itemize}

\subsubsection{Performance of open-access models}
Based on the extensive comparison of Sections \ref{sec:benchmark_accuracy}--\ref{sec:benchmark_time}, it can be concluded that the models based on DL are more likely to outperform those based on statistical methods. This is especially true in the context of  DL ensemble models as the ensemble of DNNs obtains results that are statistically significantly better than any other model. 

However, while DNNs outperformed the LEAR models, the latter are still the state-of-the-art in terms of low complexity and computational cost. In particular, their performance is very close to that of DNNs, but with the advantage of having computational costs that are up to 100 times lower. As such, they are the best available option when decision making has to be done within seconds. 

In short, new models for EPF should either be compared against LEAR models or DNNs depending on the decision time that is available. For a method to be considered more accurate than state-of-the-art methods, it should either be more accurate than the DNN model, or more accurate than LEAR but with similar or  lower computational requirements.

\section{Checklist to ensure adequate EPF research}
\label{sec:check}
As a final contribution, and with the goal of facilitating the work of reviewers of future EPF publications, we provide a short checklist to evaluate whether any new research in EPF satisfies the requirements to be reproducible and to lead to meaningful conclusions:
\begin{itemize}
	\item The test dataset comprises at least a year of data.
	\item Any new model is tested against state-of-the-art open-access models, e.g.\ the ones provided here.  
	\item The computational cost of new methods is evaluated and compared against the computational cost of existing methods.
	\item The employed datasets are open-access.
	\item The study is based on multiple markets.
	\item \rmae{} is employed as one of the accuracy metrics to evaluate forecasting accuracy.
	\item Statistical testing is used to assess whether differences in performance are significant.
	\item Forecasting models are recalibrated on a daily basis and not simply estimated once and evaluated in the full out-of-sample dataset.
	\item Hyperparameters are estimated using a validation dataset that is different from the test dataset.
	\item The split and  dates  of the dataset are explicitly stated.
	\item All the inputs of the model are explicitly defined.
	\item The test dataset is selected as the last section of the full dataset and does not contain any overlapping data with the training or validation datasets.
	\item State-of-the-art and free toolboxes are used for modeling the benchmark models.
\end{itemize}

\noindent  While this is just a very short summary of the guidelines described in Section \ref{sec:guidelines}, we think it is very useful to have them summarized together for  quick evaluations of new research.

\section{Conclusion}
\label{sec:conc}
In this paper, we have derived a set of best practices for performing research in \textit{electricity price forecasting (EPF)}. Particularly, as the field of EPF lacks a rigorous approach to compare and to evaluate new forecasting models, we have analyzed different factors affecting the quality of the research, e.g.\ dataset size or accuracy metrics, and we have proposed solutions to ensure that new research is adequate, reproducible, and useful.

In addition, as comparisons in EPF are often done using datasets that no other researches has ever tested, we have proposed an extensive open-access benchmark dataset comprising 6 years of recent data in 5 different markets. The aim of the benchmark dataset is to provide a common framework for future research so that new methods can be validated under the same conditions and meaningful comparisons can be obtained. To facilite future research, we have developed an open-source \texttt{python} library named \texttt{epftoolbox} \cite{benchmarkwebsite,epftoolboxdoc} that provides easy access to these datasets.

Similarly, as new methods in EPF are often not compared with well-established methods, we have proposed several  state-of-the-art open-source models based on statistical methods and deep learning. The methods are  tuned automatically and require no expert knowledge in order to be used. These methods are provided as open-source within the proposed \texttt{epftoolbox} library \cite{benchmarkwebsite,epftoolboxdoc} so that other researches can employ them as benchmarks in their own studies. Although the proposed methods are currently developed in \texttt{python},  we would like to extend the support to other languages; in that spirit, we encourage other researchers to help us do so.

Finally, to have a complete open-access benchmark, we have evaluated the two proposed open-access methods in the open-access dataset and we have provided the results in terms of accuracy metrics and statistical testing. Using these results, we have shown that deep neural networks  are more likely to outperform LEAR methods but that the latter are the best model for applications with short decision timeframes. Moreover, we have also shown that ensemble methods often obtain significantly better results than their individual counterparts. Based on the same results, we have also showed the importance of the guidelines as to what constitutes good practices. The most notable guidelines were that MAPE is an unreliable metric that should be avoided, that statistical testing is mandatory to obtain meaningful conclusions, and that the length of the test dataset should be at least one year.

\section*{Acknowledgment}
This research has received funding from the European Union's Horizon 2020 research and innovation program under the Marie Skłodowska-Curie grant agreement No.\ 675318 (INCITE), the Ministry of Science and Higher Education (MNiSW, Poland) through grant No.\ 0219/DIA/2019/48 and the National Science Center (NCN, Poland) through grant No.\ 2018/30/A/HS4/00444.


\bibliography{bibtex/bibtex}

\end{document}